\documentclass[journal,draftcls,onecolumn,12pt,twoside]{IEEEtran}
\normalsize

\usepackage{graphicx}
\usepackage{rotating}
\usepackage{tikz}
\usepackage{amssymb}
\usepackage{epsfig}
\usepackage{subfig}
\usepackage{epstopdf}
\usepackage{algorithm}
\usepackage{algorithmic}
\usepackage{amsmath}
\usepackage{amsfonts}
\usepackage{dsfont}
\usepackage{url}
\usepackage{chemarr}
\usepackage{chemarrow}
\usepackage{bbold}
\usepackage[version=3]{mhchem}
\usepackage{mathabx}
\usepackage{tabularx,booktabs}
\usepackage{booktabs}
\usepackage{amsmath}

\newif\ifusetables
\usetablesfalse

\ifCLASSINFOpdf
\else
\fi

\begin{document}
\title{Improving the capacity of molecular communication using enzymatic reaction cycles}
%
%
%

\author{Hamdan Awan,~\IEEEmembership{Student Member,~IEEE,}
        and Chun Tung Chou~\IEEEmembership{Member,~IEEE,}
\thanks{H. Awan and C. T. Chou are with the School
of Computer Science and Engineering, The University of New South Wales, Sydney,
NSW 2052, Australia. e-mail: (hawan,ctchou@cse.unsw.edu.au)}}
%
%

%
%

\markboth{}%
{}
%



\maketitle
\begin{abstract}
This paper considers the capacity of a diffusion-based molecular communication link assuming the receiver uses chemical reactions. The key contribution is we show that enzymatic reaction cycles, which is a class of chemical reactions commonly found in cells consisting of a forward and a backward enzymatic reaction, can improve the capacity of the communication link. The technical difficulty in analysing enzymatic reaction cycles is that their reaction rates are nonlinear. We deal with this by assuming that the amount of certain chemicals in the enzymatic reaction cycle is large. In order to simplify the problem further, we use singular perturbation to study a particular operating regime of the enzymatic reaction cycles. This allows us to derive a closed-form expression of the channel gain. This expression suggests that we can improve the channel gain by increasing the total amount of substrate in the enzymatic reaction cycle. By using numerical calculations, we show that the effect of the enzymatic reaction cycle is to increase the channel gain and to reduce the noise, which results in a better signal-to-noise ratio and in turn a higher communication capacity. Furthermore, we show that we can increase the capacity by increasing the  total amount of substrate in the enzymatic reaction cycle. 
\end{abstract}


%
\IEEEpeerreviewmaketitle

\section{Introduction}
\label{sec:intro} 
Molecular communication is a promising approach to realize communication between nano-scale devices \cite{Akyildiz:2008vt,Hiyama:2010jf,Nakano:2014fq} and especially the internet of bio-nano things \cite{akyildiz2015internet}. Molecular communication has many useful applications such as detection of harmful pathogens in environment and detection of tumour cells in human bodies \cite{nakano2012molecular}. A key characteristic of molecular communication is the use of molecules as the information or signal carrier. The transmission of signalling molecules can be carried out by diffusion \cite{Pierobon:2010kz} or active transport \cite{farsad2011simple}. In this paper, we consider diffusion-based molecular communication and in particular its information transmission capacity.

The information capacity of a communication link sets a fundamental limit on its communication performance \cite{gallager1968information}. The information capacity of molecular communication has been studied in a number of papers. We can divide these papers into two categories depending on whether they consider reactions at the receiver. The first category of articles, e.g. \cite{Atakan:2010bj,Einolghozati:2011cj,Pierobon:2013cl}, assumes a  receiver that can count the number of molecules within the receiver volume. These articles do not consider the reactions at the receiver. The second category of articles considers various types of chemical reactions at the receiver. The typical reaction considered is ligand-receptor binding, e.g. in \cite{Einolghozati:2011ge,Thomas:dc,aminian2015capacity}. Our earlier work \cite{Chou:2014jca} considers a few different types of reactions at the receiver, including linearised form of ligand-receptor binding, catalysis and regulated catalysis. This paper will focus on the capacity of chemical reaction based receivers. 

Researchers in biology have found that certain networks of chemical reactions, which are referred to as motifs, appear more often than the others in gene regulatory networks and protein reaction networks \cite{Alon}. These motifs can be considered to be the basic modules (or building blocks) to realise cell functions. Recently, there is a growing interest in the engineering and synthetic biology communities to interconnect these modules to create artificial molecular circuits \cite{Nilgiriwala:2015jj,DelVecchio:2008gy}. To be best of our knowledge, there appears to be few work on studying how the interconnection of modules will impact on the communication performance. In our earlier work \cite{Awan:2016:RER:2967446.2967455}, we study the impact on the capacity of a molecular communication link by a module consisting of a forward and a backward linearised catalytic reaction. In this paper, we consider enzymatic reaction cycles (ERCs) of the form:
\begin{align}
\cee{
K + Z &<=> KZ  -> K + Z_* \label{re:erc1} \\
P + Z_* &<=> PZ_*   -> P + Z 	\label{re:erc2} 
   }
\end{align}
where $K$ and $P$ are the enzymes catalysing the conversion between substrates $Z$ and $Z_*$, and {\color{black} $KZ$ (resp. $PZ_*$) is a complex\footnote{The reader may refer to the online entry of Compendium of Chemical Terminology (published by the International Union of Pure and Applied Chemistry) for a detailed definition of the term complex \url{http://goldbook.iupac.org/html/C/C01203.html}.} formed by the binding of $K$ and $Z$ (resp. $P$ and $Z_*$) molecules.} These ERCs are commonly found in cells and perform covalent modifications to proteins; they cover reactions such as phorsphorylation, methylation and acetylation \cite{Alberts}. The key contribution of this paper is to show that it is possible to use ERCs to improve the communication capacity of diffusion-based molecular communication. We do this by combining ERCs with receiver molecular circuits that we have studied earlier in \cite{Chou:2014jca} and show that the combinations with ERCs have a higher capacity than those without. A technical difficulty with studying ERCs is that their reaction rates are nonlinear. We assume the amount of certain {\color{black} chemical} species  {\color{black} in the ERC} is large and appeal to the tool of singular perturbation and consider a particular operation regime. This allows us to derive a closed-form expression of the channel gain. This expression suggests that we can improve the channel gain by increasing the total amount of substrate in the ERC. By using numerical calculations, we show that the effect of the ERC is to increase the channel gain and to reduce the noise, which results in a better signal-to-noise ratio and in turn a higher communication capacity. Furthermore, we show that we can improve the system capacity by increasing the total amount of substrate.

We organize this paper as follows. Section \ref{sec:related} presents related work. Section \ref{Sec:3} summarizes the modelling and analysis framework. 
In Section \ref{Sec:4}, we present the ERC and how singular perturbation can be used to derive the channel gain. Section \ref{Sec:5} presents numerical results on channel gain, noise and system capacity. Finally, Section \ref{Sec:6} concludes the paper.

\section{Related work} 
\label{sec:related}
The interest of research community in molecular communication is on the rise as shown by recent surveys \cite{Akyildiz:2008vt,Hiyama:2010jf,nakano2012molecular,Nakano:2014fq,farsad2016comprehensive}. 

The main components of a molecular communication system are transmitter, propagation medium and receiver. On the transmitter side different modulation schemes have been proposed in literature such as molecule shift keying, frequency shift keying, pulse position modulation, concentration shift keying and reaction shift keying \cite{ShahMohammadian:2012iu,Kuran:2011tg,Atakan:2010bj,Mahfuz:2011te,7208820,Awan:2015:IRM:2800795.2800798,Awan17}. 

For the propagation medium different models have been used in molecular communication literature. For example the papers \cite{Mahfuz:2011kg,mosayebi2014receivers} assume that medium is continuous while in this paper, as well as in our previous works \cite{7208820,Chou:rdmex_tnb,Chou:rdmex_nc}, we assume that the medium is divided into voxels. The use of voxels provides a convenient way to integrate diffusion and reactions into one mathematical model, see \cite{Erban:2007we} for a tutorial introduction and our earlier work for more details. An alternative end-to-end model appears in \cite{Pierobon:2011ve,Pierobon:2011vr} which is based on particle tracking. 

For the receiver side, different receiver designs have been proposed in literature for molecular communication systems, e.g. \cite{Noel:2014fv,Noel:2014hu,Kilinc:2013by,Mahfuz:2014vs,Meng:2014hh}. Similarly different demodulation techniques for molecular communication systems are presented in \cite{Noel:2014hu,Mahfuz:2014vs,awan2016demodulation,Chou:gc}.  An alternative way of designing receivers for molecular communication is by using molecular circuits, see \cite{Chou:2014jca,Chou:hf,Chou:2012ug} for example. The effect of different receiver molecular circuits on the  communication performance has been studied in the literature.  For a molecular communication system the key performance parameters are noise and capacity. The noise properties of ligand-receptor binding type of receivers are studied in \cite{Pierobon:2014iu,Pierobon:2011ve}. The  information theoretic analysis on capacity of molecular communication system is discussed in \cite{Mahdavifar15}. The information transmission capacity of different types of receiver molecular circuits is compared in \cite{Chou:2014jca}. The capacity analysis for molecular communication based on ligand receptor binding has been presented in \cite{Einolghozati:2011ge,Einolghozati:2011cj,Thomas:dc}.

One significant area of research is to study and improve the capacity of molecular communication system. For this we start with  understanding the behaviour of connecting modules in molecular communication system similar to that in synthetic biological systems \cite{cardinale2012contextualizing}. This concept of modularity in cell biology and signal transduction networks is discussed in \cite{del2013control, del2008modular,saez2004modular}. However, there appears to be few work in molecular communication where the receiver is realised by interconnecting modules. In our previous work \cite{Awan:2016:RER:2967446.2967455}, we presented an idea to improve the capacity of the molecular communication link by introducing a module consisting of simplified (i.e. linearised) ERCs. In this paper, we consider ERCs with nonlinear reaction rates which are harder to analyse due to nonlinearities. 

{\color{black}
In the existing literature, there are a number of papers which study how enzyme can be used to improve the performance of molecular communication networks. For instance, the papers \cite{Noel:2014fv} \cite{yilmaz2016interference, cho2017effective} use enzymes to reduce the amount of inter-symbol interference and the paper \cite{chude2014diffusion} studies a molecular communication link whose receiver uses an enzyme to produce output molecules from signalling molecules. They key difference between our work and these papers is that we use an ERC which consists of both forward and backward catalytic reactions while the other papers use only forward catalytic reactions. 
}

\section{Modelling and Analysis Framework}
\label{Sec:3}
\begin{figure}[t]
\begin{center}
\subfloat[The OM-only configuration.]{\includegraphics[width=10cm]{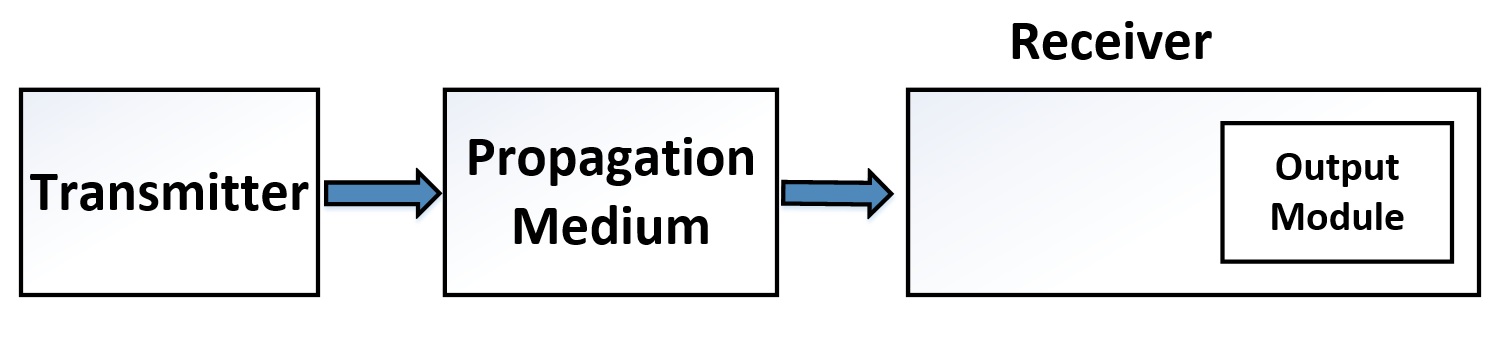} \label{fig:OM-only}} \\
\subfloat[The ERC-OM configuration.]{\includegraphics[width=10cm]{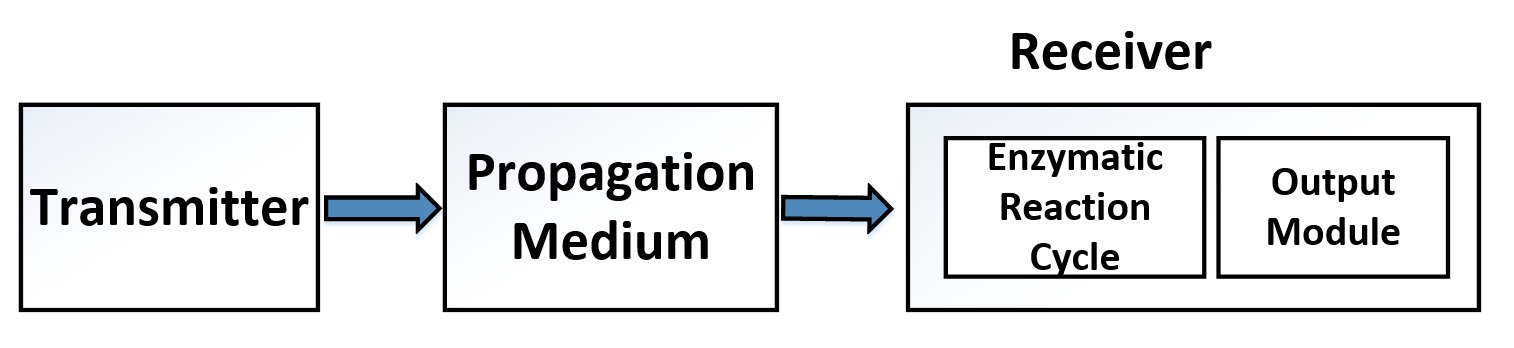} \label{fig:ERC-OM}}
 \caption{Two different link configurations considered in this paper.}
 \label{fig:config} 
 \end{center}
\end{figure}
{\color{black} We consider a molecular communication link which consists of a transmitter and a receiver.} The transmitter emits signalling molecules which diffuse freely in the propagation medium. When the signalling molecules reach the receiver, they react with the chemical reactions within the receiver to produce output molecules. The number of output molecules in the receiver over time is the output signal of the communication link. Our aim is to maximise the communication capacity of the link. 

{\color{black} In this paper, we will consider two different receiver configurations, see Figure \ref{fig:config}, where each receiver configuration is composed of different modules and each module is composed of a number chemical reactions.} In the configuration in Figure \ref{fig:OM-only}, the receiver consists only of an output module; we will refer to this as the OM-only link where OM is short for output module. In the configuration in Figure \ref{fig:ERC-OM}, the receiver consists of an ERC followed by an output module. We will refer to this as the ERC-OM link. A key contribution of this paper is to show that the ERC-OM link has a higher communication capacity than the OM-only system. 

{\color{black}
This section is organised as follows. We describe how the transmitter and propagation medium are modelled in Subsections \ref{sec:medium} and \ref{23}. This is followed by the modelling of the output module in Subsection \ref{3d}. The models for transmitter, propagation medium and output module are then combined in Subsection \ref{complete} to give a model for the OM-only system. Subsection \ref{111} explains how the capacity the OM-only system can be computed. 
}

\subsection{Propagation Medium and Transmitter} 
\label{sec:medium} 
\subsubsection{Propagation Medium}
\begin{figure}
 \begin{center}
 \includegraphics[width=9cm]{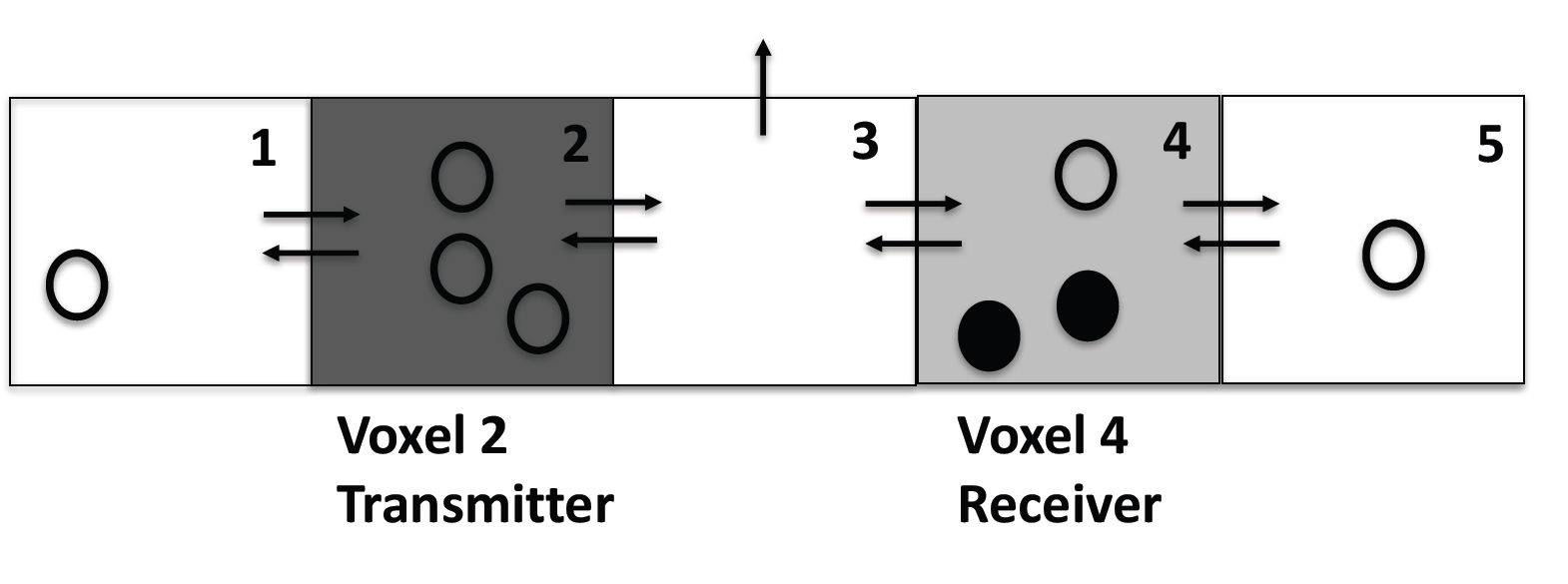}
 \caption{The propagation medium}
 \label{as:1}
 \end{center}
 \end{figure}

We assume the medium of propagation is a three dimensional space of dimension $\ell_X \times \ell_Y \times \ell_Z$ where each dimension is an integral multiple of length $ \Delta$ i.e. there exist positive integers $M_x$, $M_y$ and $M_z$ such that $\ell_X = M_x \Delta$, $\ell_Y = M_y\Delta$ and $\ell_Z = M_z \Delta $. The medium is divided into $M_x \times M_y \times M_z$ cubic voxels where the volume of each voxel is $ \Delta^3$. Figure \ref{as:1} shows an example with $M_x$ = 5 and $M_y$ = $M_z$ = 1. We assume that each voxel is given a unique index. The indices of the voxels are given in the top-right corner of the voxels in Figure \ref{as:1}. 

Diffusion is modelled by molecules moving from one voxel to a neighbouring voxel. The arrows in Figure \ref{as:1} show the directions of the movement of the molecules. We assume that the medium is homogeneous with the diffusion coefficient for the signalling molecule in the medium is $D$. Define $d = \frac{D}{\Delta^2}$. The diffusion of molecules from a voxel to a neighbouring voxel takes place at a rate of $d$, i.e. within an infinitesimal time $\delta t$, the probability that a molecule diffuses to a neighbouring voxel is $d \delta t$. 

The set up in Figure \ref{as:1} can be used to realise both reflecting and absorbing boundary conditions. A reflecting boundary condition means the molecules do not leave the medium. An absorbing boundary condition means once a signalling molecule leaves the medium, it will never return. An example of absorbing boundary condition is shown in Figure \ref{as:1} where signalling molecules can leave a surface of Voxel 3 at an escape rate of $e$.

We assume the transmitter and the receiver each occupies a voxel. However, it is straightforward to generalise to the case where a transmitter or a receiver occupies multiple voxels. The transmitter and receiver are assumed to be located, respectively, at the voxels with indices $T$ and $R$. For example, in Figure \ref{as:1}, Voxel 2 (dark grey) contains the transmitter and Voxel 4 (light grey) contains the receiver. Hence $T = 2$ and $R = 4$ for this example. 

\subsubsection{Transmitter}
The transmitter emits signalling molecules (denoted by $L$) at a rate of $u(t)$ at time $t$. This means that in the time interval [$t$, $t+\delta t$), the transmitter emits $u(t)\delta t$ signalling molecules into the transmitter voxel. We consider $u(t)$ as the input signal of the molecular communication link. We further assume that $u(t) = c + w(t)$ , where $c$ is a positive constant and $w(t)$ is {\color{black} a zero-mean stationary Gaussian random process}. Later on, we will use the spectral property of the signal $w(t)$ to maximise the mutual information through water filling \cite{gallager1968information}. 


\subsection{Diffusion-only Subsystem}
\label{23}
In this section we consider the molecular communication link in Figure \ref{as:1} assuming that the receiver reaction mechanism has been removed. {\color{black} This means the system consists of the transmitter which injects signalling molecules into the transmitter voxel and the diffusion of signalling molecules in the propagation medium.} We will refer to this as the diffusion-only subsystem and it is the same for both the OM-only and ERC-OM links. Our aim is to illustrate how this subsystem can be modelled. Although the example is based on the network in Figure \ref{as:1}, generalisation to other networks is straightforward. 

The state of the diffusion-only subsystem is the number of signalling molecules in each voxel. Let $n_{L,i}(t)$ denote the number of signalling molecules in Voxel $i$, then the state of the diffusion-only subsystem in Figure \ref{as:1} is given by:
\begin{align}
n_L(t) =
\left[ \begin{array}{ccccc}
 n_{L,1}(t)&n_{L,2}(t)&n_{L,3}(t)& n_{L,4}(t)&n_{L,5}(t)
\end{array} \right]^T
\label{1}
\end{align}
where the superscript $^T$ denotes matrix transpose. 

Our basic modelling framework is to consider each movement of a signalling molecule from a voxel to another as an event. For example let us consider the movement of a molecule from Voxel 2 to Voxel 3. This movement occurs at a rate of $dn_{L,2}$ and after this movement has taken place, $n_{L,2}$ will be decreased by 1 and $n_{L,3}$ increased by 1. We can describe the change in the number of molecules in the voxels due to this movement by a jump vector. For this example, the jump vector $q_{d} = [0,-1,1,0,0]^T$ and the state changes from $n_{L}(t)$ to $n_{L}(t) + q_{d}$ due to this movement. The rate of this movement can be represented by a jump rate $W_{d}(n_L(t))$ $= d n_{L,2}$. We can use exactly the same concept to model the boundary condition. 

In general, we write the jump vectors $q_{d,j}$ and jump rates $W_{d,j}(n_L(t))$ where the subscript $d$ is used to indicate an event due to diffusion and $j$ is used to index the events. We use $J_d$ to denote the total number of diffusion events. For Figure \ref{as:1}, $J_d = 9$ consisting of 8 inter-voxel diffusion events and 1 escape event. With the 9 jump vectors and jump rates, we can use stochastic differential equation (SDE) \cite{gardiner2009stochastic} to model the dynamics of the diffusion-only subsystem. {\color{black} This modelling framework is based on the two facts: first, diffusion between voxels can be modelled as abstract chemical reactions \cite{Erban:2007we}; second, chemical reactions can be modelled by SDE \cite{Higham:2008dl}.} The diffusion of signalling molecules in the configuration in Figure \ref{as:1} can be modelled by the following SDE:          
\begin{align}
\dot{n}_L(t) & = \sum_{j = 1}^{J_d} q_{d,j}W_{d,j} (n_L(t)) + \sum_{j = 1}^{J_d} q_{d,j} \sqrt{W_{d,j}(n_L(t))} \gamma_j + {\mathds 1}_T u(t)
\label{eqn:sde:do} 
\end{align}
{{\color{black}  where $\gamma_j$ is continuous Gaussian white noise with unit variance and ${\mathds 1}_T$ is a unit vector with 1 at the $T$-th element with the subscript $T$ being the index of the transmitter.} 
There are three terms on the right-hand side of Eq.~\eqref{eqn:sde:do} and we will discuss them one by one. The first term describes the deterministic dynamics. Since all the jump rates of all the diffusion events are linear, this term can be written as a product of a matrix $H$ and the state vector $n(t)$. The matrix $H$ is defined by the following equality: 
\begin{align}
H n_L(t) & = \sum_{j = 1}^{J_d} q_{d,j}W_{d,j} (n_L(t))  
\end{align}
For Figure \ref{as:1}, the $H$ matrix is given by: 
\begin{align}
H =  
\left[ \begin{array}{ccccc}
-d & d & 0 & 0 & 0 \\
d & -2d & d & 0 & 0 \\
0 & d & -{\color{black} 2}d-e & d  & 0 \\
0 & 0 & d & -2d & d \\
0 & 0 & 0 & d  & -d \\
\end{array} \right]
\label{eqn:H} 
\end{align}
The second term of Eq.~\eqref{eqn:sde:do} describes the stochastic dynamics. {\color{black} An intuitive way to understand the first two terms in Eq.~\eqref{eqn:sde:do} is as follows. Over a finite time interval $\Delta t$, the number of times that the $j$-th type of jump occurs can be approximated by a Poisson random variable with mean $W_{d,j} (n_L(t)) \; \Delta t$. If $W_{d,j} (n_L(t)) \; \Delta t$ is large, then we know from probability theory that a Poisson random variable with mean $W_{d,j} (n_L(t)) \; \Delta t$ can be approximated by a Gaussian variable with both mean and variance given by $W_{d,j} (n_L(t)) \; \Delta t$. We can therefore approximate the number of times that the $j$-th type of jumps occurs by $W_{d,j} (n_L(t)) \; \Delta t + \sqrt{W_{d,j} (n_L(t)) \; \Delta t} \gamma_j$ where the first term is the mean number of jumps and the second term is the deviation from the mean; these two terms give rise to the first two terms in Eq.~\eqref{eqn:sde:do}. For a more detailed explanation, the reader can refer to \cite{Higham:2008dl}.} 

The third term models the transmitter. Since the transmitter emits $u(t)\delta t$ molecules at time $t$, we add this number of molecules to voxel $T$ (the index of the transmitter voxel) at time $t$. 

It is important to point out that the elements in $n_{L}(t)$, which have the interpretation of the number of molecules, is strictly speaking a {\sl discrete} random variable. The SDE is an approximation which holds when the order of the number of molecules is ${\cal O}(100)$ \cite{deRonde:2012fs}. However, as far as the first and second order moments are concerned, the SDE \eqref{eqn:sde:do} gives the same result as a master equation formulation that assumes the number molecules is discrete \cite{warren2006exact}.


\subsection{Output Module}
\label{3d}
The output signal of both OM-only and ERC-OM links is the count of the number of output molecules over time. The aim of the output module is to produce the output molecules, and hence the output signal. In our earlier work \cite{Chou:2014jca}, we studied the impact of a number of molecular circuits on the capacity of the communication link. We will use two of the molecular circuits in \cite{Chou:2014jca} as output modules in this paper. Our aim is to show that the improvement of capacity by ERCs is general and applies to multiple types of output modules.  

Each output module has two chemical species $B$ and $X$ where $X$ is the output molecule. The identity of $B$ depends on whether we are using OM-only or ERC-OM links. For the case of OM-only links, the chemical species $B$ is the signalling molecules in the receiver voxel. We will explain what the identity of $B$ is for the ERC-OM case later on. We assume that the output molecules $X$ do not diffuse and stay inside the receiver voxel. 

Following the terminology of \cite{Chou:2014jca}, the output modules are referred to as reversible conversion (RC) and catalysis plus regulation (CARTEG). 


\subsubsection{The RC output module}
\label{rc}
The RC output module consists of two linear chemical reactions and they are described by their chemical reaction equation, jump vector and jump rate. 
\begin{align}
B & \rightarrow X 		& \left[ \begin{array}{cc} -1 & 1 \end{array} \right]^T&, k_+ n_{B}  \label{cr:rc1}  \\
X & \rightarrow B		& \left[ \begin{array}{cc} 1 & -1 \end{array} \right]^T&, k_- n_X       \label{cr:rc2}
\end{align}
where $n_B$ and $n_X$ are respectively the number of $B$ and $X$ molecules, and $k_+$ and $k_-$ are reaction rate constants. In reaction \eqref{cr:rc1}, the jump vector $[-1 \; 1]$ indicates that a $B$ molecule is converted to an $X$ molecule in this reaction while the reaction \eqref{cr:rc2} is the reverse of this, hence the name reversible conversion. 


\subsubsection{The CATREG output module}
\label{catreg}
The CATREG output module consists of three chemical reactions mentioned below with their respective respective jump functions and jump rates.
\begin{align}
B &  \rightarrow  B +  X 		& \left[ \begin{array}{cc} 0 & 1 \end{array} \right]^T&, k_+n_{B}  \label{cr:rc12}  \\
X & \rightarrow \phi		& \left[ \begin{array}{cc} 0 & -1 \end{array} \right]^T&, k_- n_X       \label{cr:rc22}
\\
B & \rightarrow_X \phi		& \left[ \begin{array}{cc} -1 & 0 \end{array} \right]^T&, k_0 n_X       \label{cr:rc23}
\end{align}

In reaction \eqref{cr:rc12} the chemical $B$ acts as a catalyst to produce the output molecule X at a rate of $k_+n_{B}$. Note that  the number of $B$ molecules remains unchanged before and after the reaction. This is indicated by the jump vector  which shows that every time this reaction occurs, the number of $B$ remains unchanged and the number of output molecules is increased by $1$. The  reaction \eqref{cr:rc22} is a degradation reaction where $X$ molecules are degraded to a species $\phi$ that we are not interested to keep track of. Lastly in reaction \eqref{cr:rc23}  the degradation of $B$ molecules in the receiver voxel is driven by the presence of the output molecules $X$ at a rate of $k_0 n_X$ and we can view this as using $X$ to regulate the amount of $B$. 
 
\subsubsection{Modelling the output module}
\label{sec:model:om}
\label{225}
This section presents an SDE model of the output module. This model will be used for the modelling of the OM-only and ERC-OM links later on. The output module consists of two chemicals. The state vector of the output module is:
\begin{align}
 \tilde{n}_R(t) 
 & =  \left[ \begin{array}{c|c}
 n_{B}(t) & n_X(t)  
\end{array} \right]^T 
\label{eq:state_ntildeR}
\end{align} 

We will use $q_{r,j}$ and $W_{r,j}(\tilde{n}_R(t)$ to denote, respectively, the jump vector and jump rates in the output module. Note that the subscript $r$ indicates the jump vector and jump rate belong to the output module, which is situated inside the receiver. Let $J_r$ be the number of reactions in the output module; the values of $J_r$ for RC and CATREG are, respectively, 2 and 3. We will index $j$ from $J_d+1$ to $J_d+J_r$ so that we can keep the expression simple when we combine the diffusion-only subsystem and the output module later. By using the jump vectors and jump rates of the output module, the SDE that describes the evolution of the number of molecules in the output module is: 
\begin{align}
\dot{\tilde{n}}_R(t) & = {\cal R} \tilde{n}_R(t) + \sum_{j = J_d+1}^{J_d+J_r} q_{r,j} \sqrt{W_{r,j}(\tilde{n}_R(t))} \gamma_j 
\label{eqn:sde:ro11} 
\end{align}
where $\gamma_j$ is continuous {\color{black} Gaussian} white noise of unit variance and the ${\cal R}$ matrix is defined by the relation: $ {\cal R} \tilde{n}_R(t) = \sum_{j =  J_d +1}^{J_d + J_r} q_{r,j} W_{r,j}( \tilde{n}_R(t) )$. The ${\cal R}$ matrices for the RC and CATREG output modules are shown in Table \ref{table:1}. {\color{black} The derivation of Eq.~\ref{eqn:sde:ro11} follows from the fact that the dynamics of chemical reactions can be approximately modelled by SDE \cite{Higham:2008dl}. The modelling technique being used is similar to that in deriving Eq.~\ref{eqn:sde:do}.}


\begin{table}[]
\centering
\caption{${\cal R}$ matrix for different output modules}
\begin{tabular}{|c|c|}
\hline
\multicolumn{1}{|c|}{Output module}	&	\multicolumn{1}{|c|}{${\cal R}$ Matrix}	\\
\hline
RC &    $ \begin{bmatrix}  -k_{+} & k_{-} \\ k_{+} & -k_{-} \end{bmatrix}$
\\ \hline
CATREG      & $ \begin{bmatrix} 0 & - k_{0}  \\ k_{+} & -k_{-} \end{bmatrix}$ 
\\ \hline
\end{tabular}

\label{table:1}
\end{table}

\subsection{The OM-only link}
\label{complete}
In this section, we will combine the SDE models for the diffusion-only subsystem in Section \ref{23} and the output module in Section \ref{3d} to form the complete model for the OM-only link. It is important to note that in the OM-only link, the $B$ molecule in the OM is the signalling molecules in the receiver voxel. Therefore, the interconnection between the diffusion-only subsystem and the output module is the number of signalling molecules in the receiver voxel, which is common to both of them. 

{\color{black}
We will use the example in Fig.~\ref{as:1} to explain how the diffusion-only subsystem and the output module can be combined together. The dynamics of the diffusion-only subsystem for  Fig.~\ref{as:1} is given by Eq.~\ref{eqn:sde:do}. For this example, the receiver voxel has the index $R = 4$, so the evolution of the number of signalling molecules in the receiver voxel $n_{L,R}(t)$ is given by the $R$-th (i.e. fourth) row of Eq.~\ref{eqn:sde:do}, which is: 
\begin{align}
\dot{n}_{L,R}(t) = d n_{L,3}(t) - 2d n_{L,R}(t) + d n_{L,5}(t) + \underbrace{\sum_{j = 1}^{J_d} [q_{d,j}]_R \sqrt{W_{d,j}(n_L(t))} \gamma_j}_{\xi_d(t)}
\label{eqn:sde:ro:r2}  
\end{align}
where $[q_{d,j}]_R$ denote the $R$-th element of the vector $q_{d,j}$. 
}

{\color{black}
In the output module, the signalling molecules in the receiver voxel are the $B$ molecules in Section \ref{3d}. The dynamics of the number of signalling molecules in the receiver voxel due to the reactions in the output module is given by the first element of Eq.~\ref{eqn:sde:ro11}, which is: 
\begin{align}
\dot{n}_{L,R}(t) = R_{11} n_{L,R}(t) + R_{12} n_X(t) + \underbrace{\sum_{j = J_d+1}^{J_d+J_r} [q_{r,j}]_1 \sqrt{W_{r,j}(\tilde{n}_R(t))} \gamma_j}_{\xi_r(t)} 
\label{eqn:sde:ro:r1} 
\end{align}
where $[q_{r,j}]_1$ denotes the first element of the vector $q_{r,j}$. 
} 

For the OM-only link, the dynamics of $n_{L,R}(t)$ is obtained by combining Eq. \eqref{eqn:sde:ro:r2} and \eqref{eqn:sde:ro:r1} as follows: 
\begin{align}
\dot{n}_{L,R}(t) = &  d n_{L,3}(t) - 2d n_{L,R}(t) + d n_{L,5}(t)   + 
R_{11} n_{L,R}(t)  + R_{12} n_X(t) + \xi_{total}(t) 
\label{eqn:sde:nlr:ex} 
\end{align}
where $\xi_{total}(t) = \xi_d(t) + \xi_r(t)$.  

We are now ready to describe the complete model for the OM-only link. Let $n(t)$ be the state of the OM-only link and it is given by: 
\begin{align}
n(t) = 
 & \left[ \begin{array}{c|c}
 n_{L}(t)^T & n_X(t)  
\end{array} \right]^T
\label{eqn:state} 
\end{align}

We will also need to modify the jump vectors from the diffusion-only subsystem and the output module to obtain the jump vectors for the complete model; this will be explained in a moment.  We use $q_j$ and $W_j(n(t))$ to denote the jump vectors and jump rates of the combined model. The SDE for the complete system is:
\begin{align}
\dot{n}(t) & = A n(t) + \sum_{i = 1}^{J} q_j \sqrt{W_j(n(t))} \gamma_j + {\mathds 1}_T u(t) 
\label{eqn:mas11}
\end{align} 
where $J = J_d+J_r$, and the matrix $A$ is defined by $A n(t) = \sum_{i = 1}^{J} q_j W_j(n(t))$. The matrix $A$ has the block structure: 
\begin{align}
A = 
 & \left[ \begin{array}{c|c}
H + {\mathds 1}_R^T {\mathds 1}_R R_{11}  &   {\mathds 1}_R R_{12}  \\ \hline 
R_{21}  {\mathds 1}_R^T & R_{22} 
\end{array} \right]
\label{eqn:A} 
\end{align}
where $H$ comes from the diffusion only subsystem (Note: an example of $H$ for Figure \ref{as:1} is in \eqref{eqn:H}.) and $R_{11}$, $R_{12}$ etc come from the output module. The vector ${\mathds 1}_R$ is a unit vector with an 1 at the $R$-th position; in particular, note that  ${\mathds 1}_R^T n_L(t) = n_{L,R}(t)$ which is the number of signalling molecules in the receiver voxel. Note that, the coupling between the diffusion-only subsystem and the output module, as exemplified by \eqref{eqn:sde:nlr:ex}, takes place at the $R$-th row of $A$. 

We now explain how the jump vectors for the OM-only link are formed. Let $m_d$ denote the dimension of the vector $n_L(t)$. Note that $m_d$ is in fact the number of voxels. The dimension of the jump vectors $q_j$ in the complete system is $m_d+1$. Given jump vector $q_{d,j}$ ($j = 1,...,J_d$) from the diffusion only subsystem with dimension $m_d$, we append a zero to $q_{d,j}$ to obtain $q_j$. The jump vectors $q_{r,j}$ ($j = J_d+1,...,J_d+J_r$) from the output module has dimension $m_r+1$. To obtain $q_j$ from $q_{r,j}$, we do the following: (1) take the first element of $q_{r,j}$ and put it in the $R$-th element of $q_j$; (2) take the last element of $q_{r,j}$ and put it in the last element of $q_j$. Note that jump rates are unchanged when combining the subsystems.

\subsection{Capacity of the OM-only link}
\label{111}
The input and output signals for the OM-only link are, respectively, the production rate $u(t)$ of the signalling molecules in the transmitter voxel and the number of output molecules $n_X(t)$ in the receiver voxel. In this section, we will derive an expression for the mutual information between the input $u(t)$ and output $n_X(t)$. 

{\color{black} 
We begin by stating a result in \cite{Tostevin:2010bo} which states that, for two Gaussian distribution random processes $a(t)$ and $b(t)$, their mutual information $I(a,b)$ is given by: 
\begin{align}
I(a,b) &= \frac{-1}{4\pi} \int_{-\infty}^{\infty} \log \left( 1 - \frac{|\Phi_{ab}(\omega)|^2}{\Phi_{aa}(\omega) \Phi_{bb}(\omega)}  \right) d\omega
\label{eqn:MI0}
\end{align} 
where $\Phi_{aa}(\omega)$ (resp. $\Phi_{bb}(\omega)$) is the power spectral density of $a(t)$ ($b(t)$), and $\Phi_{ab}(\omega)$ is the cross spectral density of $a(t)$ and $b(t)$.} 

{\color{black}
In order to apply the above results to the communication link given in Eq.~\eqref{eqn:mas11}, we need a result from \cite{warren2006exact} on the power spectral density of systems consisting only of chemical reactions with linear reaction rates. Following from \cite{warren2006exact}, if all the jump rates $W_j(n(t))$ in \eqref{eqn:mas11} are linear in $n(t)$, then the power spectral density of $n(t)$ is given by the following SDE:
\begin{align}
\dot{n}(t) & = A n(t) + \sum_{i = 1}^{J} q_j \sqrt{W_j(\langle n(\infty) \rangle)} \gamma_j + {\mathds 1}_T u(t) 
\label{eqn:complete2}
\end{align} 
where $\langle n(t) \rangle)$ denotes the mean of $n(t)$ and is the solution to the following ordinary differential equation:
\begin{align}
\dot{\langle n(t) \rangle} & = A \langle n(t) \rangle+ {\mathds 1}_T c 
\label{eqn:ode_ninfinity}
\end{align} 
where $c$, which was defined before, is the mean of input $u(t)$.  
}

{\color{black}
As a result, the dynamics of the OM-only link in Eq.~\eqref{eqn:complete2} are described by a set of {\sl linear} SDE with $u(t)$ as the input and $n_X(t)$ (which is the last element of the state vector $n(t)$) as the output. The input $u(t)$ has the form $u(t) = c + w(t)$ where $c$ is a constant to set the operating point of the system and $w(t)$ is a zero-mean Gaussian random process. The noise in the output $n_X(t)$ is caused by the Gaussian white noise $\gamma_j$'s in Eq.~\eqref{eqn:complete2}. Therefore, Eq.~\eqref{eqn:complete2} models a continuous-time linear time-invariant (LTI) stochastic system subject to Gaussian input and Gaussian noise. 
}

The power spectral density $\Phi_{X}(\omega)$ of the signal $n_X(t)$ can be obtained from standard results on the output response of a LTI system to a stationary input \cite{Papoulis} and is given by: 
\begin{align}
\Phi_{{X}}(\omega) & =  |\Psi(\omega) |^2 \Phi_u(\omega) + \Phi_{\eta}(\omega)
\label{2331} 
\end{align}
where $\Phi_u(\omega)$ is the power spectral density of $u(t)$ and $|\Psi(\omega)|^2$ is the channel gain with $\Psi(\omega) = \Psi(s)|_{s = i\omega}$ defined by:
\begin{align}
\langle N_{X} (s) \rangle  & = {\mathds 1}_X \langle N(s)  \rangle = 
\underbrace{ {\mathds 1}_X  (sI - A)^{-1} {\mathds 1}_T }_{\Psi(s) }   U(s)
\label{21}
\end{align}
Note that Eq.~\eqref{21} can be obtained from Eq.~\eqref{eqn:complete2} after taking the mean and applying Laplace transform. The term $\Phi_{\eta}(\omega)$ denotes the stationary noise spectrum and is given by: 
\begin{align}
\Phi_{\eta}(\omega) & =   \sum_{j = 1}^{J_d + J_r} | {\mathds 1}_X (i \omega I - A)^{-1} q_j |^2 W_j(\langle n_{}(\infty) \rangle) 
\label{eqn:spec:noise2} 
\end{align} 
where $n_{}(t)$ denotes the state of the complete system in \eqref{eqn:state} and $\langle n_{}(\infty) \rangle$ is the mean state of system at time $\infty$ due to constant input $c$. 

{\color{black}
Similarly, by using standard results on the LTI system, the cross spectral density $\Psi_{xu}(\omega)$ has the following property:
\begin{align}
|\Psi_{xu}(\omega)|^2 &= |\Psi(\omega) |^2 \Phi_u(\omega)^2 
\label{eqn:csd} 
\end{align} 
} 

{\color{black}
By substituting Eq.~\eqref{2331} and Eq.~\eqref{eqn:csd} in the mutual information expression in Eq.~\eqref{eqn:MI0}, we arrive at the mutual information $I(n_{X},u)$ between $u(t)$ and $n_{X}(t)$ is:
\begin{align}
I(n_{X},u) = \frac{1}{2} \int \log \left( 1+\frac{ | \Psi(\omega) |^2}{\Phi_{\eta}(\omega)} \Phi_u(\omega) \right) d\omega
\label{eqn:mi}
\end{align}
The capacity of the link can be determined by applying the water-filling solution to \eqref{eqn:mi} subject to power constraint on input $u(t)$ \cite{gallager1968information}. The capacity of the link depends on the channel gain, noise power spectral density and the input power spectral density. 
}

\section{The ERC-OM link}
\label{Sec:4}
\label{sec:erc} 
The aim of this section is to use analytical methods to study the property of the ERC-OM link. We learn from Section \ref{111} that the channel gain can be used to influence the capacity of the communication link. However, channel gain is only defined for LTI systems but the reaction rates of the ERC are nonlinear functions of the concentration of the reactants. In order to study the channel gain of the ERC-OM link, we use singular perturbation and assume the amount of certain chemical species is large to obtain a linear approximation of the input-output response of a ERC-OM link. We derive closed-form expressions for the approximate channel gain for the ERC-OM link with RC or CATREG as the output module. An insight from these closed-form expressions is that we can increase the channel gain by increasing the amount of certain chemical species in the ERC module. We will show using numerical studies in Section \ref{sec:eval} that these chemical species can be used to increase the capacity of the ERC-OM link. 

\subsection{Enzymatic Reaction Cycles (ERC)}
\label{ERC}
The ERC consists of two sets of chemical reactions which facilitate the conversion of a pair of chemical species. We will refer to the pair of chemicals as $Z$ and $Z_*$. We will refer to the conversion of $Z$ to $Z_*$ as forward and the conversion of $Z_*$ to $Z$ as backward. We have already shown the chemical equations for ERC in Section \ref{sec:intro} but we will rewrite them in a slightly different form so that we can simplify the notation of some equations later on. The chemical equations for the ERC are:
\begin{align}
\cee{
K + Z <=>[\beta_1][\beta_2] C_1& ->[k_1] K + Z_* \label{77} \\
P + Z_* <=>[\alpha_1][\alpha_2] C_2&  ->[k_2] P + Z 	
   \label{78}}
\end{align}
where the enzymes $K$ and $P$ catalyze, respectively, the forward and reverse conversions. The chemical species $C_1$ and $C_2$ are intermediate complexes form by the enzymes and the substrates. The symbols $k_1$, $k_2$, $\beta_1$, $\beta_2$, $\alpha_1$ and $\alpha_2$ are reaction rate constants. The enzyme $P$ can exist on its own or as part of the complex $C_2$. We use $P_T$ to denote the total amount of $P$ and we assume $P_T$ is a constant. 



\subsection{ERC with the RC output module} 
\label{sec:b}
The aim of this section is to derive the channel gain for the ERC-OM link where the output module is RC. In this setup, we assume: (1) The chemical $K$ in the ERC in reaction \eqref{77} is the signalling molecules in the receiver voxel; and, (2) The chemical $B$ in the RC reaction is $Z_*$. We also assume that all the chemicals of the ERC and output module, with the exception of signalling molecules, stay within the receiver voxel. 

The output module of this link consists of the following reactions: 
\begin{align}
\cee{
Z_* &<=>[k_+][k_-] X 
\label{cr:rc}}
\end{align}
where $X$, as before, represents the output molecule. Note that these reactions are identical to those in Section \ref{rc} except that we have replaced $B$ by $Z_*$. {\color{black} We will use $Z_T$ to denote the total amount of $Z$, $Z_*$, $C_1$, $C_2$ and $X$; and we will assume that $Z_T$ is a constant.}

Since our aim is to derive the channel gain, which relates the mean input and output signals, we will therefore use only the deterministic part of the SDEs. In order to simplify the notation, we will drop the angle brackets $\langle \; \rangle$ and all the chemical species counts are assumed to be their mean value. In addition, for the chemical species in the ERC, we will simply use their chemical names as their count. 

The dynamics of the chemical species in the ERC and output modules are: 
\begin{align}
& \dot  C_{1}(t)  =  - (\beta_2 + k_1) C_1(t)+  \beta_1 n_{L,R}(t) ( Z_T - Z_{*}(t) - C_1(t) -C_2(t) - n_X (t))\label{eq:30}  \\
& \dot  C_{2}(t)  =  - (\alpha_2 + k_2) C_2(t) +  \alpha_1  Z_{*}(t) (P_T - C_2(t)) \label{eq:30a} \\
 & \dot Z_{*} (t)  =  k_1C_{1}(t) + \alpha_2 C_2(t)  - \alpha_1  Z_{*}(t) (P_T - C_2(t)) - k_+ Z_{*} (t)   + k_-  n_{X} (t) \label{eq:30b}  \\
 & \dot n_{X} (t)  = k_+ Z_{*} (t)  - k_-  n_{X} (t)
\end{align}
Note that Eq.~\eqref{eq:30}, \eqref{eq:30a} and \eqref{eq:30b} are nonlinear. We first assume that $Z_T$  $\gg$ ($Z_*(t)$ - $C_1(t)$ - $C_2(t)$ - $n_X(t))$, so we can simplify Eq.~\eqref{eq:30} to:
\begin{align}
\dot  C_{1}(t)  =  - (\beta_2 + k_1) C_1(t) +  \beta_1 n_{L,R}(t) Z_T 
\end{align}
We next assume that $P_T \gg C_2(t)$, so we can replace the term $(P_T - C_2(t))$ in Eq.~\eqref{eq:30a} and \eqref{eq:30b} by $P_T$. As a result of these two assumptions, we have got rid of the nonlinearities in the model. 

Even after getting rid of nonlinearities, the resulting equations are still difficult to analyse. The next step is to use time scale separation between the rate of diffusion and chemical reactions to simplify the equations. We define $G_1 = \frac{ \beta_1 Z_T}{d}$ and $G_2 = \frac{k_-}{k_1}$. We assume $\epsilon_1$= $\frac{1}{G_1}$ and $\epsilon_2$= $\frac{1}{G_2}$ are small; {\color{black} we will discuss these assumptions further at the end of this sub-section.} This allows us to define the slow variable $W (t)$:
\begin{align}
W (t)  = Z_{*} (t)  +  n_{X} (t)
\label{w}
\end{align}
and apply singular perturbation to separate the fast and slow dynamics \cite{del2013control}. The essential idea behind time scale separation is that the fast dynamics reaches equilibrium before the slow dynamics. This means we can replace a differential equation describing fast dynamics by an algebraic equation and we only have to keep track of the slow dynamics. {\color{black} Note that in singular perturbation, the approximation error in replacing a differential equation describing the fast dynamics by an algebraic equation is of the order ${\cal O}(\max(\epsilon_1,\epsilon_2))$ if $\epsilon_1$ and $\epsilon_2$ are sufficiently small, see \cite{khalil2002nonlinear}. }  

%
%

After applying singular perturbation and going through many steps we obtain one algebraic equation and one ordinary differential equation (ODE), as follows: 
\begin{align}
{n_{X}}(t) &\approx  r Z_{*}(t)  \\
\dot Z_{*} (t)  &=  [G_1 k_1' C_1(t) + G_1 a_2 C_2(t) - G_1 a_1 Z_{*}(t) ] [\frac{1}{(1+r)}] 
\label{eqn:125}
\end{align}
where $k_1'$ = $\frac{k_1}{G_1}$ , $a_1$ = $\frac{\alpha_1 P_T}{G_1}$, $a_2$ = $\frac{\alpha_2}{G_1}$ and $r = \frac{k_+}{k_-}$. By using Laplace transform, we can combine these two equations as:  
%
\begin{align}
& N_{X} (s)   = r[\frac {(k_1 C_1(s)+ \alpha_2 C_2(s)/(1+r)} {s + \alpha_1 P_T/(1+r)}]  
\label{eqn:3b5}
\end{align}
By using the ODEs that describe the diffusion of the signalling molecules, we can derive the expressions of $C_1(s)$ and $C_2(s)$ in terms of the input signal $U(s)$.

{
\color{black} The expressions for $C_1(s)$ and $C_2(s)$ can be found in Eqs. \eqref{eqn:c1} and \eqref{eqn:c2}. Furthermore the complete derivation for $N_{X}(s)$ is presented in Appendix \ref{app:exp}.}


The final result is: 
\begin{align}
& N_{X}(s)   =  \underbrace {Q(s)\frac{k_1 \beta_1 Z_T}{(s + \beta_2 + k_1)}} _{\tilde{\Psi}(s)} U(s)
\label{eqn:3b6}
\end{align}
where
\begin{align}
& Q(s) = {r [\frac {( 1 + \alpha_1P_T / [(s+\alpha_2+k_2)(s+\alpha_1P_T)-(\alpha_1\alpha_2P_T)]) / (1+r)}{s + \alpha_1 P_T/(1+r)}]} {\mathds 1}_R^T (sI - H)^{-1}{\mathds 1}_T  
\label{eqn:3b7}
\end{align}
Note that the the channel gain is given by $|\tilde{\Psi}(s)|^2$ where $\tilde{\Psi}(s)$ in given in Eq.~\eqref{eqn:3b6}. It can be seen from the expression of $\tilde{\Psi}(s)$ that channel gain is an increasing function of $Z_T$. 

{
\color{black}
{\bf Discussions:} (1) In order to derive the results in this section, we assume that $Z_T$ and $P_T$ are large, and limit ourselves to the operating regime where $\epsilon_1 = \frac{d}{\beta_1 Z_T}$ and $\epsilon_2=\frac{k_1}{k_-}$ are small. If $Z_T$ is large, then $\epsilon_1$ can be made small. Therefore, the requirements we need are large $Z_T$, $P_T$ and small $k_1$ relative to $k_-$.  If the output receiver is to be constructed by synthetically, then the quantities $Z_T$ and $Y_T$ can be chosen by the designer. The condition on $k_1$ can be met by choosing an appropriate ERC. Similar methods have been used in \cite{Mishra:2014da} to realise a synthetic biomolecular circuit with certain properties. The paper \cite{Mishra:2014da} assumes that the quantities of certain chemical species are large and requires certain reaction rate constant to be large. (2) In this paper, we have limited ourselves to a particular operating regimes. It may be possible to arrive at the same result by consider other operating regimes. We will leave this for further work. 

}




\subsection{ERC with the CATREG output module}
\label{sec:c}
The aim of this section is to derive the channel gain for the ERC-OM link where the output module is CATREG. In this setup, we assume: (1) The chemical $K$ in the ERC in Reaction \eqref{77} is the signalling molecules in the receiver voxel; and, (2) The chemical $B$ in the CATREG reaction is $Z_*$. The latter assumption means that the output module of this link consists of the following reactions: 
\begin{align}
\cee{
 Z_* & ->[k_+]  Z_* + X  \label{79}\\	
    X & ->[k_-]  \phi \label{80} \\	
Z_* & ->[k_0]_X  \phi 
\label{cr:rc5}}
\end{align}
Note that these reactions are identical to those in Section \ref{catreg} except that we have replaced $B$ by $Z_*$. {\color{black} We will use $Z_T$ to denote the total amount of $Z$, $Z_*$, $C_1$ and $C_2$. Note that due to the degradation of $Z_*$ by $X$ in reaction \eqref{cr:rc5}, the value of $Z_T$ decreases with time. In this paper, we will assume that the degradation rate $k_0$ is small so $Z_T$ can be considered to be a constant over a small finite time interval.}

The dynamics of the chemical species in the ERC and output modules are: 
\begin{align}
& \dot  C_{1}(t)  =  - (\beta_2 + k_1) C_1(t) +  \beta_1 n_{L,R}(t) ( Z_T - Z_{*}(t) - C_1(t) -C_2(t))
\label{eq:43}\\
 & \dot  C_{2}(t)  =  - (\alpha_2 + k_2) C_2(t) +  \alpha_1  Z_{*}(t) (P_T-C_2(t)) \label{eq:erc-catreg:2} \\
 & \dot Z_{*} (t)  =  k_1 C_1(t) + \alpha_2 C_2(t)  - \alpha_1  Z_{*}(t) (P_T-C_2(t))  - k_0  n_{X} (t) \label{eq:erc-catreg:3} \\
 & \dot n_{X} (t)  = k_+ Z_{*} (t)  - k_-  n_{X} (t)
\end{align}
We will assume $Z_T$ $\gg $ ($Z_*(t)$ - $C_1(t)$ - $C_2(t)$) and $P_T \gg C_2(t)$ to remove the nonlinearities in the model. 
%
%
We further assume that $\epsilon_1$= $\frac{1}{G_1}$ and $\epsilon_2$= $\frac{1}{G_2}$ are small, and this allows us to define a slow variable $W(t)$ in Eq.~\eqref{w}. After applying singular perturbation and  going through numerous steps, we obtain an algebraic equation and an ODE:  
\begin{align}
{n_{X}} (t)  &\approx r Z_{*}(t) \\ 
\dot Z_{*} (t)  &= [G_1 k_1' C_1(t) + G_1 a_2 C_2(t) - G_1 a_1 Z_{*}(t)- r(k_0 + k_-)]  [1/(1+r)] 
\label{eqn:12} 
\end{align}
where $k_1'$ = $\frac{k_1}{G_1}$ , $a_1$ = $\frac{\alpha_1 P_T}{G_1}$, $a_2$ = $\frac{\alpha_2}{G_1}$, $r = \frac{k_+}{k_-}$, and the {
\color{black} expressions of $C_1(s)$ and $C_2(s)$ are given in Eq.~\eqref{eqn:c1} and \eqref{eqn:3bdss} in Appendix \ref{app:exp}.} By combining these two equations with those that describe the diffusion dynamics, we can show that $N_X(s)$ again has the form as Eq.~\eqref{eqn:3b6} but the expression of Q(s) is: 

%
%
%
%
%
%
%
%
%

\begin{align}
& Q (s) = {r [\frac {( 1 + \alpha_1P_T / [(s+\alpha_2+k_2)(s+\alpha_1P_T+ k_+ - rk_0)-(\alpha_1\alpha_2P_T)]) / (1+r)}{s + \alpha_1 P_T/(1+r)}]} {\mathds 1}_R^T (sI - H)^{-1}{\mathds 1}_T 
\label{eqn:3b9}
\end{align}
This shows that the channel gain is again an increasing function of $Z_T$. We will show in Section \ref{sec:eval} that we can use $Z_T$ to improve the capacity of the ERC-OM link when CATREG is the output module.

\subsection{Computation of capacity} 
This section explains how the capacity of an ERC-OM link can be computed. The dynamics of the link are described by the SDE:
\begin{align}
\dot{n}(t) & = \sum_{i = 1}^{J} q_j W_j( n(t) )  + \sum_{i = 1}^{J} q_j \sqrt{W_j(\langle n(t) \rangle)} \gamma_j + {\mathds 1}_T u(t) 
\label{eqn:sde:erc-om}
\end{align} 
where $q_j$'s are jump vector, $W_j(n(t))$'s are the jump rates, and $J$ is the total number of diffusion and reaction events. Note that although we have made use of the same notation as before, we assume $q_j$, $W_j(n(t))$ and $J$ have been adapted for the ERC-OM link. Since the ERC contains nonlinear reactions, it means that some of the jump rates $W_j(n(t))$ are nonlinear functions of the state vector $n(t)$. This means we cannot write the first term on the right-hand side of Eq.~\eqref{eqn:sde:erc-om} as the product of a matrix and the state vector $n(t)$.  

We argue in Sections \ref{sec:b} and \ref{sec:c} that if we assume $Z_T$ $\gg $ ($Z_*(t)$ - $C_1(t)$ - $C_2(t)$ - $n_X(t))$ and $P_T \gg C_2(t)$, then we can approximate the nonlinear dynamics with linear ones. This is the same as replacing nonlinear jump rates $W_j(n(t))$ by linear ones. Once this replacement is done, we can proceed as before to compute the mutual information and capacity.  

\section{Numerical evaluation}
\label{Sec:5}
\label{sec:eval} 

This section presents numerical examples to illustrate the use of the ERC to improve the communication performance of a molecular communication link.

{
\color{black} We assume an array of $5\times 2 \times 2$ voxels with a voxel size of ($\frac{1}{3}$$\mu$m)$^{3}$ (i.e. $\Delta = \frac{1}{3}$ $\mu$m). We have also used larger arrays of voxels and we obtain similar results. The transmitter and receiver are located at voxels (2,1,1) and (4,2,2) respectively. We assume the diffusion coefficient $D$ of the medium is $1$ $\mu$m$^2$s$^{-1}$. The deterministic emission rate $c$ is chosen to be 10 molecules per second. We assume an  absorbing boundary with an escape rate $e$ equal to $\frac{d}{10}$.  For the CATREG receiver circuit in the output module, we fix $k_0$ = $0.01$  whereas $k_+$ and $k_-$ (both or any one of them) can be varied to obtain different values of association constant $r = \frac{k_+}{k_-}$.  For the ERC module, we choose $\beta_1= \beta_2=1$, $k_1=0.05$, $\alpha_1=\alpha_2=1$ and $k_2=0.5$. Furthermore, we choose $Z_T =500$ and $P_T=200$.} 

\begin{figure}
\begin{center}
\includegraphics[trim=0cm 0cm 0cm 0cm ,clip=true, width=.65\columnwidth]{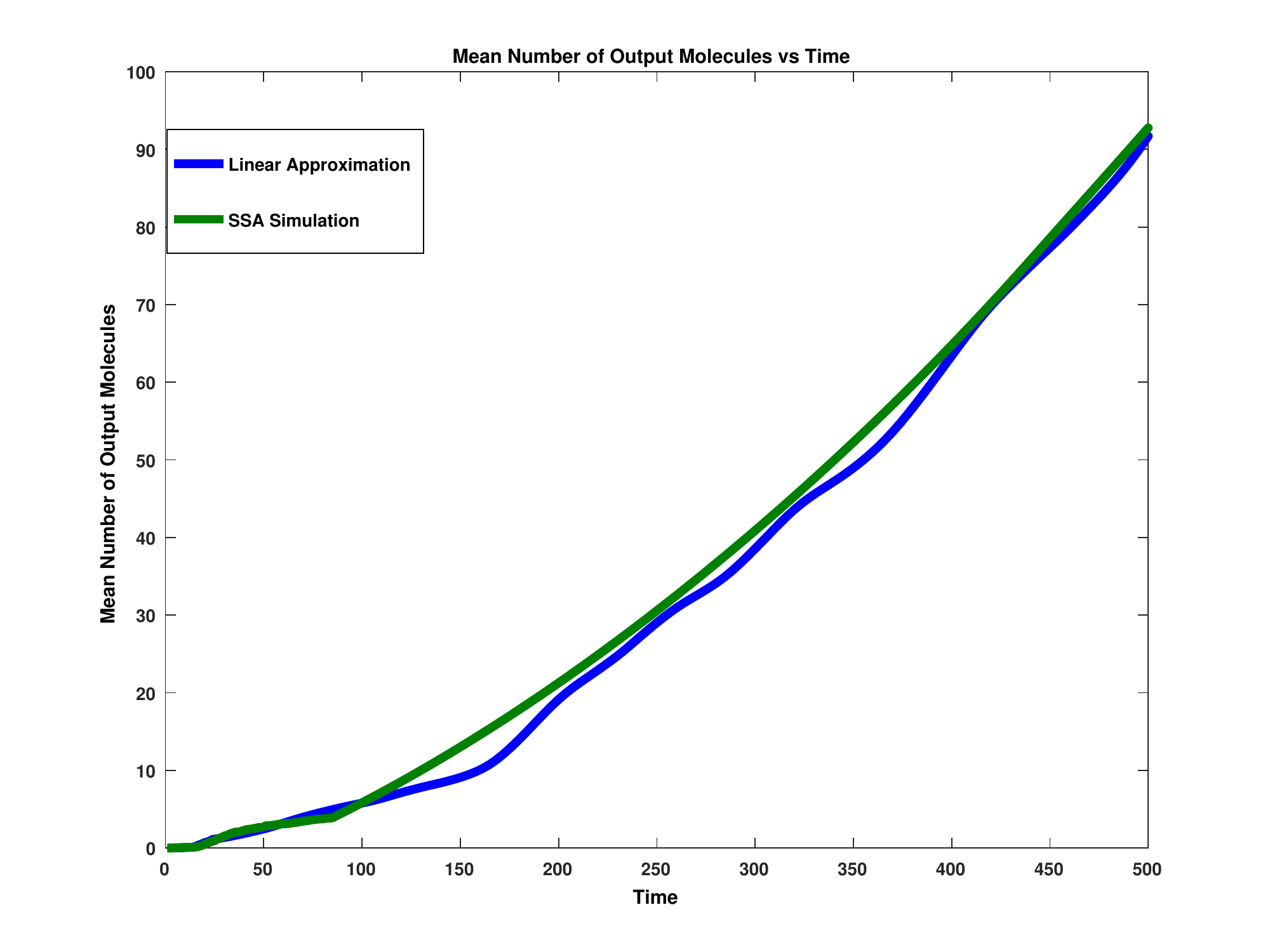}
\caption{This figure shows that SSA simulation of ERC-OM with nonlinear reaction rates gives almost the same output as the linear approximation.}
\label{16}
\end{center}
\end{figure}

The ERC-OM link is nonlinear and we derive linear approximation for the link in Sections \ref{sec:b} and \ref{sec:c}. We first verify that the linear approximation is sufficiently accurate. We do this by simulating the ERC-OM with nonlinear reaction rates and compare the results against those given by the linear approximation. For the simulation of ERC-OM with nonlinear reaction rates, we use Stochastic Simulation Algorithm (SSA) \cite{Gillespie:1996ve} which is a method to simulate systems with both chemical reactions and diffusion. The SSA algorithm simulates the Markov chain which describes the evolution of the number of molecules in a system due to diffusion and reactions \cite{Erban:2007we}. The reason why we choose SSA is because it can be applied to reactions with non-linear reaction rates. We use SSA to simulate an ERC-OM link with nonlinear reaction rates and compute the mean number of output molecules. We compare this against the mean number of output molecules given by the linear approximation. Figure \ref{16} compares the mean number of output molecules from SSA simulation and the linear approximation. It can be seen from the figure that the linear approximation is accurate.

Our next step is to show the improvement made by the ERC-OM link over that of OM-only. We have performed numerical experiments on using both RC and CATREG as the output module. We will present the results for CATREG only because the results for RC are similar. For these numerical experiments, we vary the association constant $r = \frac{k_+}{k_-}$ of the output module.

\begin{figure}
\begin{center}
\includegraphics[trim=0cm 0cm 0cm 0cm ,clip=true, width=.65\columnwidth]{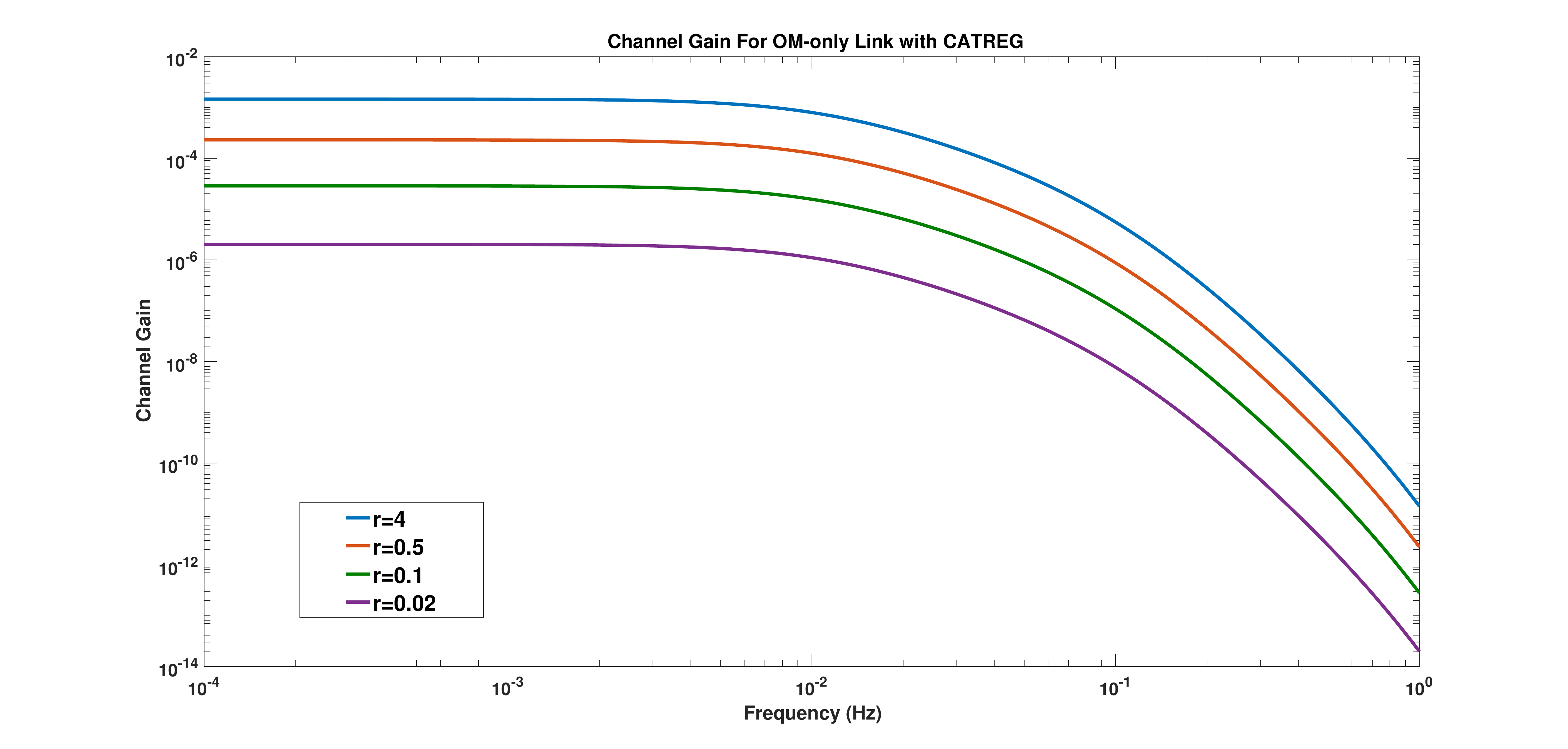}
\caption{{\color{black}The channel gain for the OM-only link with CATREG as the output module.}}
\label{10}
\end{center}
\end{figure}

\begin{figure}
\begin{center}
\includegraphics[trim=0cm 0cm 0cm 0cm ,clip=true, width=.65\columnwidth]{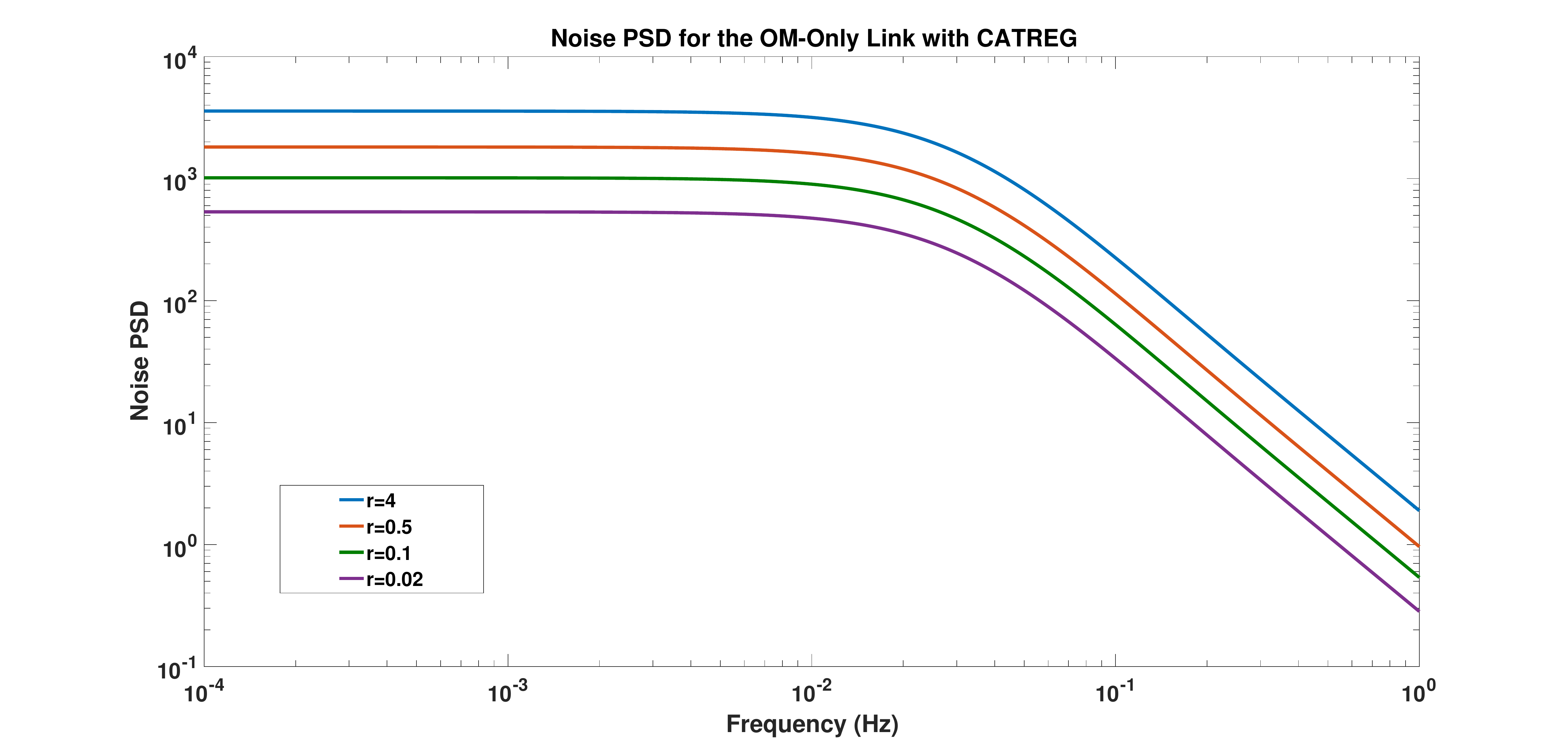}
\caption{{\color{black}The noise power spectral density for the OM-only link with CATREG as the output module.}}
\label{11}
\end{center}
\end{figure}
 \begin{figure}
 \begin{center}
 \includegraphics[trim=0cm 0cm 0cm 0cm ,clip=true, width=.65\columnwidth]{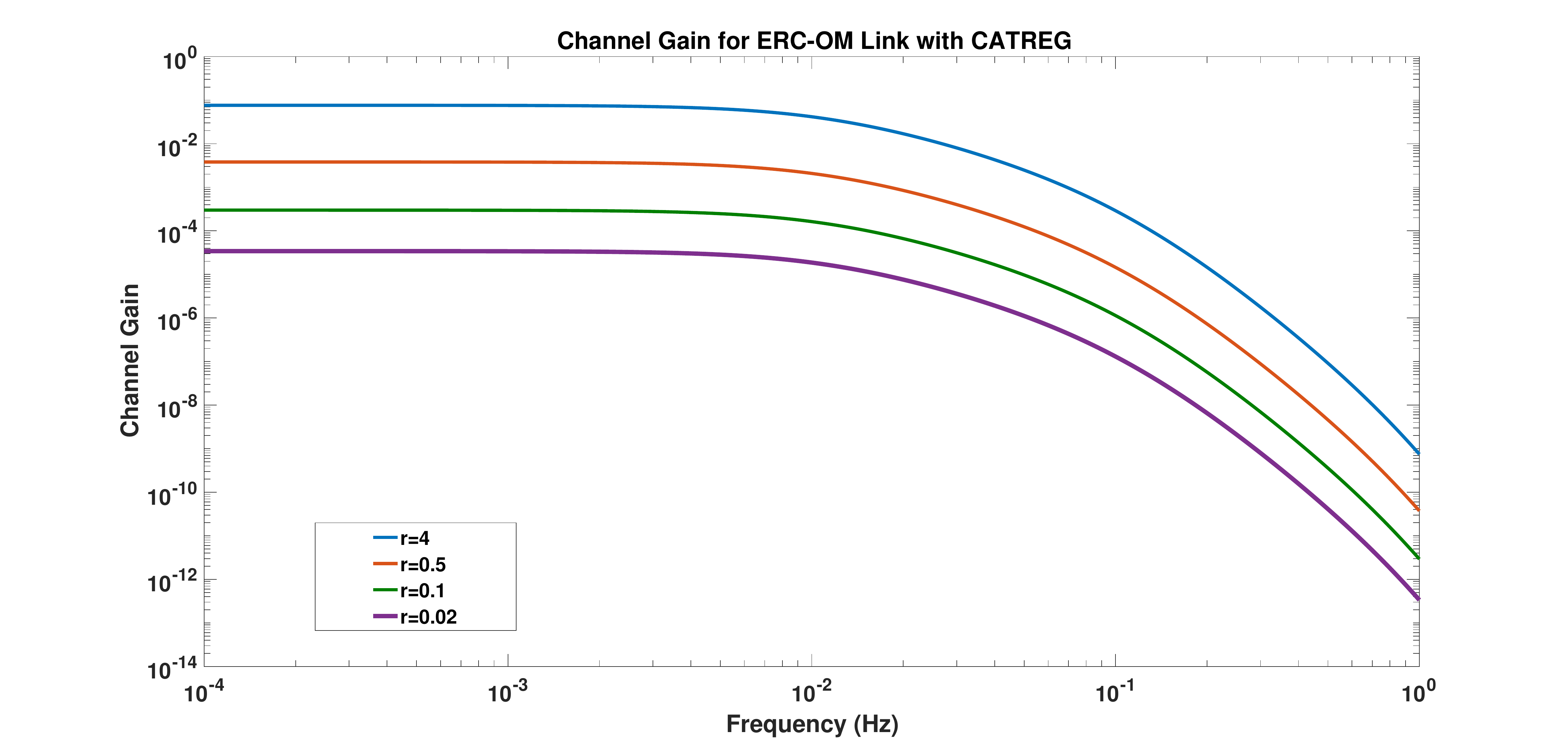}
 \caption{{\color{black}The channel gain for the ERC-OM link with CATREG as the output module.}}
 \label{12A}
 \end{center}
 \end{figure}
We first compare the channel gain and noise power spectral density of the OM-only and ERC-OM links. Figures \ref{10} and \ref{11} show, respectively, the channel gain and noise power spectral density of the OM-only link for different values of $r$. The corresponding results for ERC-OM links are shown in Figures \ref{12A} and \ref{13}. These figures are best viewed in colour because we have used the same coloured line for the same value of $r$. By comparing Figures \ref{10} and \ref{12A}, we can see that the effect of the ERC is to increase the channel gain. By comparing Figures \ref{11} and \ref{13}, we can see that the effect of the ERC is to decrease the noise power spectral density. This means the overall effect of ERC is to increase the signal-to-noise ratio of the link, and therefore capacity.

 \begin{figure}
 \begin{center}
 \includegraphics[trim=0cm 0cm 0cm 0cm ,clip=true, width=.65\columnwidth]{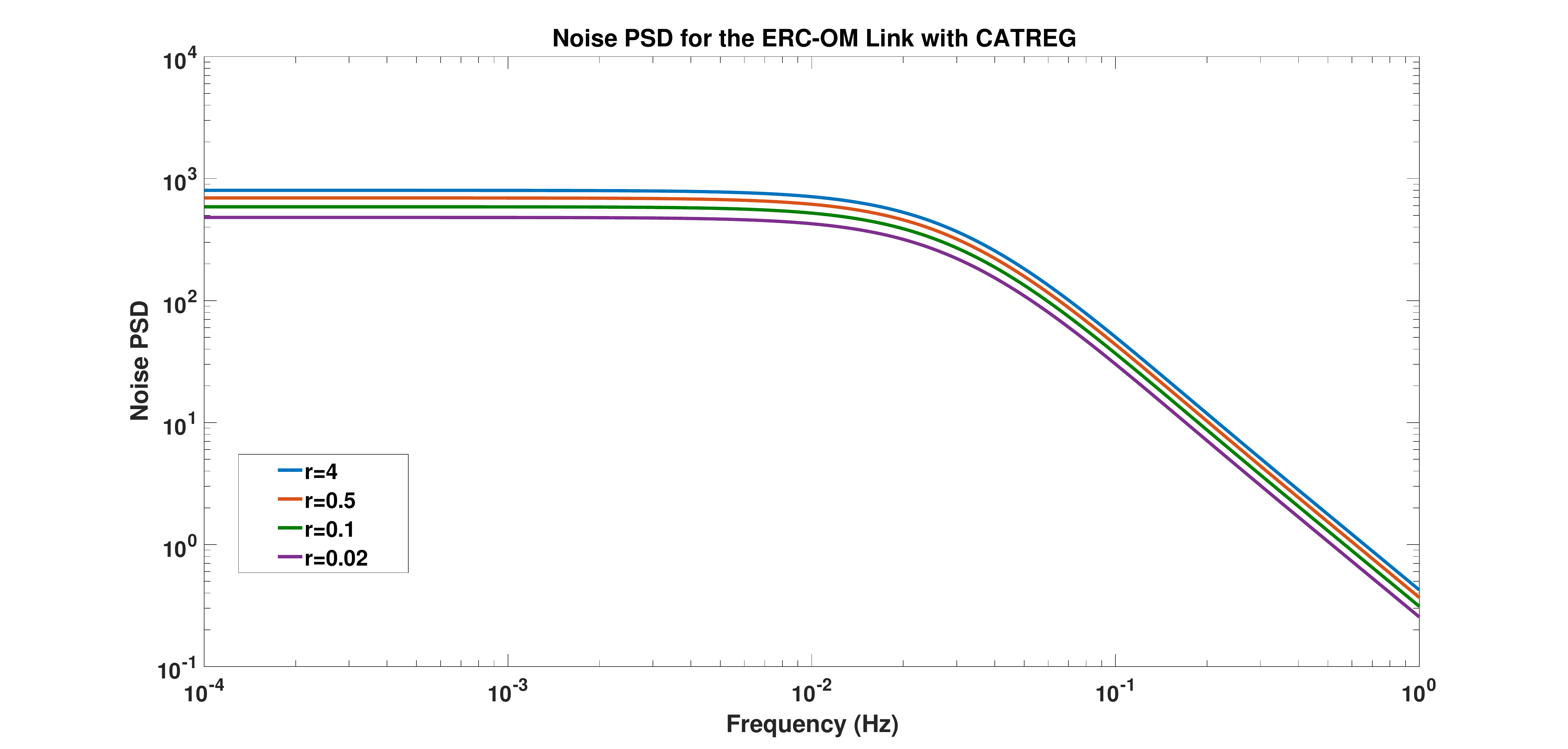}
 \caption{{\color{black}The noise power spectral density for the ERC-OM link with CATREG as the output module.}}
 \label{13}
 \end{center}
 \end{figure} 
 
We now compare the capacity of the OM-only and ERC-OM links. In these numerical experiments, we vary $k_+$ from 0 to 10 and keep other reaction rate constants unchanged. We use two different pairs of $Z_T$ and $P_T$. We compute the capacity of the OM-only and ERC-OM links. Figure \ref{15} shows the results for $Z_T = 500$ and $P_T = 200$ and Figure \ref{14} is for $Z_T = 2000$ and $P_T = 500$. We see in both Figures \ref{15} and \ref{14} that the ERC can improve the link capacity for all values of $r$. The figures also show that the capacity increases with $r$ initially but plateaus off later on. 

We next study how the value of $Z_T$ impacts on the capacity. The derivation in Sections \ref{sec:b} and \ref{sec:c} shows that $Z_T$ can increase the channel gain. We vary $Z_T$ from 500 to 5000. We use two different values of $k_+$ and keep all the other reaction rate constants unchanged. Figure \ref{a} shows that we can increase the capacity by increasing $Z_T$ and this increase is observed for both values of $k_+$ being used. This is a welcoming news because it gives us a method to increase the capacity of the link by adjusting the amount of chemical being used. If one thinks about the degrees of freedom that one can use to influence the chemical reactions in a receiver, one can change the amount of chemical species or the reaction rate constants. Unfortunately, it may not always be possible to change reaction rate constants because they depend on temperature and pressure of the operating environment which may be out of our control. However, the amount of chemical species is a parameter that can readily be controlled. Therefore, our research provides a practical method to tune the capacity of a communication link.

\begin{figure}
\begin{center}
\includegraphics[trim=0cm 0cm 0cm 0cm ,clip=true, width=.65\columnwidth]{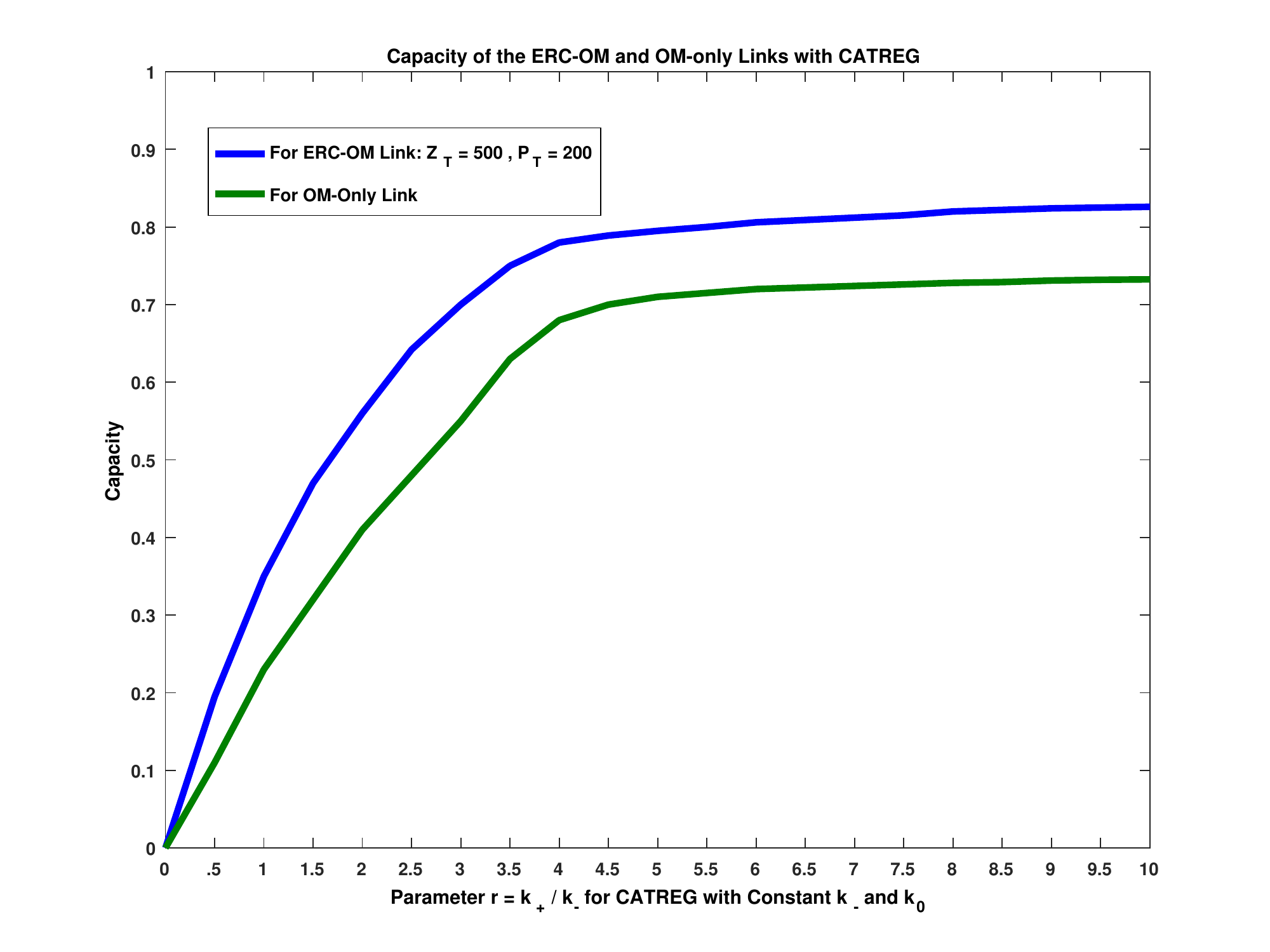}
\caption{{\color{black}Capacity of the ERC-OM and OM-only links for $Z_T = 500$ and $P_T = 200$.}}
\label{15}
\end{center}
\end{figure}

\begin{figure}
\begin{center}
\includegraphics[trim=0cm 0cm 0cm 0cm ,clip=true, width=.65\columnwidth]{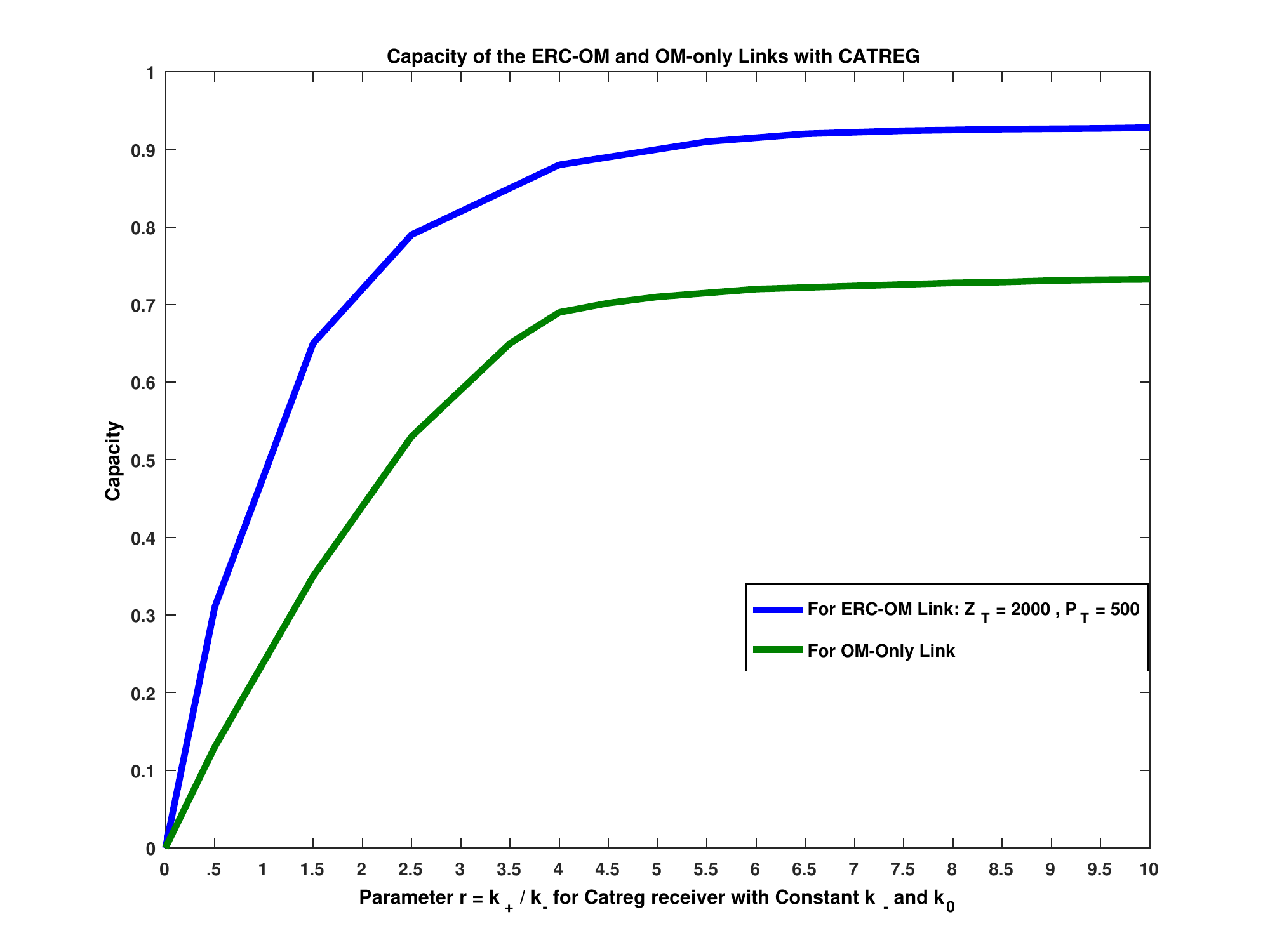}
\caption{{\color{black}Capacity of the ERC-OM and OM-only links for $Z_T = 2000$ and $P_T = 500$.}}
\label{14}
\end{center}
\end{figure}

\begin{figure}
\begin{center}
\includegraphics[trim=0cm 0cm 0cm 0cm ,clip=true, width=.65\columnwidth]{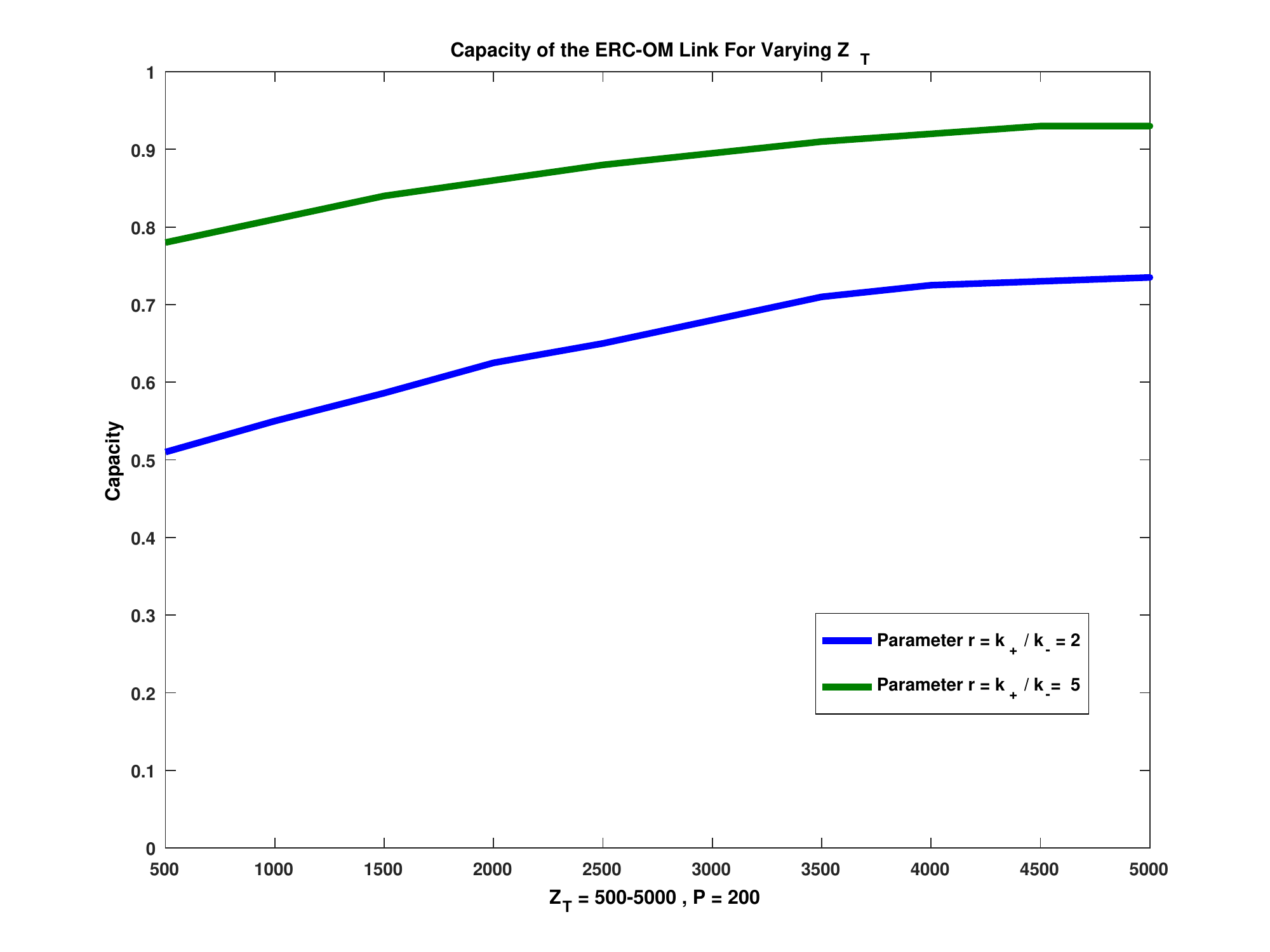}
\caption{{\color{black}Effect on capacity with varying $Z_T$.}}
\label{a}
\end{center}
\end{figure}

\section{Conclusions}
\label{Sec:6}
This paper considers the capacity of a communication link in diffusion-based molecular communication. We consider the case that the receiver uses chemical reactions. Our key contribution is that we show that enzymatic reaction cycles can be used to increase the capacity of communication link. We further show that we can increase the capacity of the link by increasing the amount of a certain chemical species in the enzymatic reaction cycles. This provides a practical way to adjust the link capacity.

\newpage

\appendices 
\section{Expressions}
\label{app:exp}

{\color{black}

Note that in this section we present the derivation of Equations \eqref{eq:30}-\eqref{eq:30b} and \eqref{eqn:3b6} for ERC-OM link with RC receiver circuit only. The equations for the ERC-OM link with CATREG receiver circuit can be derived using the same process.

For ease of reference, we first re-write the reactions in the ERC as mentioned in Section \ref{ERC}.

\begin{align}
\cee{
& K + Z <=>[\beta_1][\beta_2] C_1  \label{7a} \\
& C_1 ->[k_1] K + Z_* \label{7b } \\
& P + Z_* <=>[\alpha_1][\alpha_2] C_2	 \label{7c} \\
 & C_2  ->[k_2] P + Z    \label{7d}}
\end{align}

Furthermore we present the reactions in the output Module (i.e. RC receiver circuit) as:

\begin{align}
\cee{
Z_* &<=>[k_+][k_-] X 
\label{7e}}
\end{align}

We obtain equation for $Z_{*} (t)$  from the above equations as follows:

\begin{align}
& \dot Z_{*} (t)  =  k_1C_{1}(t) + \alpha_2 C_2(t)  - \alpha_1  Z_{*}(t) (P_T - C_2(t)) - k_+ Z_{*} (t)   + k_-  n_{X} (t) \label{eq:newa}  
\end{align}

The first term $k_1C_{1}(t)$ is obtained as  a result of Reaction \eqref{7b }. The second term $\alpha_2 C_2(t)$ is obtained as  a result of backward Reaction in \eqref{7c}. Similarly the third term is obtained as a result of forward Reaction in \eqref{7c} where the number of $P$ molecules is given by $P_T - C_2(t)$. Finally the last two terms are obtained as a result of forward and backward reactions in \eqref{7e} respectively.

Similarly, we find the expressions for $ \dot  C_{1}(t) $ and $ \dot  C_{2}(t)$ from the reactions in the ERC as follows:
\begin{align}
& \dot  C_{1}(t)  =  - \beta_2 C_1(t) - k_1 C_1(t) + \beta_1 n_{L,R}(t) ( Z_T - Z_{*}(t) - C_1(t) -C_2(t) - n_X (t))\label{eq:30v}  \\
& \dot  C_{2}(t)  =  - \alpha_2 C_2(t) - k_2 C_2(t) +  \alpha_1  Z_{*}(t) (P_T - C_2(t)) \label{eq:30av} 
\end{align}

For $ \dot  C_{1}(t) $ the first term comes from the backward Reaction in \eqref{7a} whereas the second term comes from the forward reaction in \eqref{7c}. The last term comes from the forward reaction in \eqref{7a} where the number of signalling molecules $K$ in receiver  is given by $n_{L,R}(t)$ and the number of $Z$ molecules is given by $Z_T - Z_{*}(t) - C_1(t) -C_2(t) - n_X (t)$.

Similarly for $ \dot  C_{2}(t) $ the first term comes from the backward Reaction in \eqref{7c} whereas the second term comes from the forward reaction in \eqref{7d}. The last term comes from the forward reaction in \eqref{7c} where the number of signalling molecules $Z_{*}$ in receiver  is given by $Z_{*}(t)$ and the number of $P$ molecules is given by $P_T - C_2(t)$.

Finally the forward and backward reactions in \eqref{7e} give us following equation for the number of output molecules $n_{X} (t)$: 

\begin{align}
 & \dot n_{X} (t)  = k_+ Z_{*} (t)  - k_-  n_{X} (t)
 \label{m1}
\end{align}

Note that Eq.~\eqref{eq:newa}, \eqref{eq:30v} and \eqref{eq:30av} and  are nonlinear. To remove this non-linearity we first assume that the constant $Z_T$  $\gg$ ($Z_*(t)$ - $C_1(t)$ - $C_2(t)$ - $n_X(t))$. This enables us to simplify Eq.~\eqref{eq:30v} as:
\begin{align}
\dot  C_{1}(t)  =  - (\beta_2 + k_1) C_1(t) +  \beta_1 n_{L,R}(t) Z_T 
\label{c1}
\end{align}

We next assume that constant $P_T$ $\gg$  $C_2(t)$, so we can replace the term $P_T$ -  $C_2(t)$ in Eq. \eqref{eq:30av} by $P_T$.

\begin{align}
& \dot  C_{2}(t)  =  - \alpha_2 C_2(t) - k_2 C_2(t) +  \alpha_1  Z_{*}(t) P_T  \label{eq:new} 
\end{align}

Using the same assumption  $P_T$ $\gg$  $C_2(t)$ in Eq. \eqref{eq:newa} we get:

\begin{align}
& \dot Z_{*} (t)  =  k_1C_{1}(t) + \alpha_2 C_2(t)  - \alpha_1  Z_{*}(t) P_T  - k_+ Z_{*} (t)   + k_-  n_{X} (t) \label{eq:newb}  
\end{align}

The resulting equations are still difficult to analyse so we use time scale separation between the rate of diffusion and chemical reactions for further simplification. We define $G_1 = \frac{ \beta_1 Z_T}{d}$ and $G_2 = \frac{k_-}{k_1}$. We assume $\epsilon_1$= $\frac{1}{G_1}$ and $\epsilon_2$= $\frac{1}{G_2}$ are small and define a slow variable $W (t)$:

\begin{align}
W (t)  = Z_{*} (t)  +  n_{X} (t)
\label{w1}
\end{align}

Using this value in Eq.\eqref{m1} we get:

\begin{align}
 & \dot n_{X} (t)  = k_+ Z_{*} (t)  - k_-  n_{X} (t)
 \label{m2}
\end{align}

Next we apply singular perturbation to separate the fast and slow dynamics. To do this we multiply both sides by a small $\epsilon_2 = \frac{k_1}{k_-}$ and get :

\begin{align}
 & \epsilon_2 \dot n_{X} (t)  = \epsilon_2  k_+ Z_{*} (t)  - \epsilon_2 k_-  n_{X} (t)
 \label{m3}
\end{align}

\begin{align}
 & \epsilon_2 \dot n_{X} (t)  = \frac{k_1}{k_-}  k_+ Z_{*} (t)  - \frac{k_1}{k_-} k_-  n_{X} (t)
 \label{m4}
\end{align}

Next we set $\epsilon_2$ equal to zero and simplify the equation as follows:

\begin{align}
 & 0 \times  \dot n_{X} (t)  \approx k_1 \frac{k_+}{k_-}   Z_{*} (t)  - k_1 n_{X} (t)
 \label{m5}
\end{align}


\begin{align} 
 &   n_{X} (t)  = r   Z_{*} (t)  
 \label{m7}
\end{align}

where $r=  \frac{k}{k_-}$

This means that we can simplify Equation \eqref{w1} as:

\begin{align}
W (t)  =  Z_{*} (t)  +   n_{X} (t)
\label{w2}
\end{align}

\begin{align}
 W (t)  =  Z_{*} (t)  [1 + r]
\label{w3}
\end{align}

Next we define $G_1 = \frac{ \beta_1 Z_T}{d}$, $G_2 = \frac{k_-}{d}$, $k_1' = \frac{k_1}{G_1}$, $k_d =\frac{k_-}{k_+}$, $a_1$ = $\frac{\alpha_1 P_T}{G_1}$, $a_2$ = $\frac{\alpha_2}{G_1}$ and write $Z_{*} (t) $ and  $ n_{X} (t)$ from Eq. \eqref{eq:newa} and \eqref{m1} respectively using these notations: 

\begin{align}
\dot  Z_{*} (t)  = G_1 k_1' C_1(t) + G_1 a_2 C_2(t) - G_1 a_1 Z_{*}(t)  + G_2 d n_{X} (t) - G_2 \frac{d}{k_d} Z_{*}(t)
\label{w6}
\end{align}

\begin{align}
\dot n_{X} (t) = - G_2 d n_{X} (t) + G_2 \frac{d}{k_d} Z_{*}(t)
\label{w7}
\end{align}

Next we solve for $\dot  W (t)$ as follows :

\begin{align}
\dot  W (t)  = \dot  Z_{*} (t)  + \dot n_{X} (t)
\label{w8}
\end{align}

Putting the values from Eqs.  \eqref{w6} and  \eqref{w7} we get:

\begin{align}
\dot  W (t)  = G_1 k_1' C_1(t) + G_1 a_2 C_2(t) - G_1 a_1 Z_{*}(t) 
\label{w9}
\end{align}

We also know that:

\begin{align}
\dot  W (t)  = \dot  Z_{*} (t)  [1 + r]
\label{w4}
\end{align}

Comparing both these equations we get following equation:

\begin{align}
\dot  Z_{*} (t)  [1 + r] = G_1 k_1' C_1(t) + G_1 a_2 C_2(t) - G_1 a_1 Z_{*}(t)  
\label{w5}
\end{align}

%

\begin{align}
\dot Z_{*} (t)  &=  [G_1 k_1' C_1(t) + G_1 a_2 C_2(t) - G_1 a_1 Z_{*}(t) ] [\frac{1}{(1+r)}] 
\label{eqn:125}
\end{align}

Next by using Laplace transform, we can obtain following equation:

\begin{align}
 Z_{*} (s)  &=  \frac{[G_1 k_1' C_1(s) + G_1 a_2 C_2(s)]/(1+r)}{ s +  G_1 a_1   /(1+r) }\\
 \end{align}
 
 Which reduces to following by using the values of $G_1$, $k_1'$, $a_1$ and $a_2$.
 \begin{align}
 Z_{*} (s)  & = [\frac {(k_1 C_1(s)+ \alpha_2 C_2(s)/(1+r)} {s + \alpha_1 P_T/(1+r)}]
\label{eqn:125b}
\end{align}

Furthermore we take Laplace of Eq. \eqref{m7} and put the value of $Z_{*} (s)$  from Eq. \eqref{eqn:125b} to obtain:

\begin{align}
 N_{X} (s)= r  Z_{*} (s) 
\end{align}

%
\begin{align}
& N_{X} (s)   = r[\frac {(k_1 C_1(s)+ \alpha_2 C_2(s)/(1+r)} {s + \alpha_1 P_T/(1+r)}]  
\label{eqn:3b5a}
\end{align}

The next aim is to derive expressions of $C_1(s)$ and $C_2(s)$ respectively. We perform the Laplace transform of \eqref{c1} and go through some steps to obtain $C_1(s)$ in terms of the input signal $U(s)$ as follows:

\begin{align}
& C_{1} (s) = \frac {\beta_1 Z_T} {(s + \beta_2 + k_1)} N_{L,R}(s) = \frac {\beta_1 Z_T} {(s + \beta_2 + k_1)} {\mathds 1}_R^T (sI - H)^{-1} {\mathds 1}_T U(s)
\label{eqn:c1}
\end{align}
where  $ N_{L,R}(s) = {\mathds 1}_R^T (sI - H)^{-1} {\mathds 1}_T U(s)$.

Similarly to obtain $C_2(s)$ we perform the Laplace transform of \eqref{eq:new} and go through some steps to simplify it as:
\begin{align}
& C_{2} (s) = \frac { \alpha_1 P } {(s+\alpha_2+k_2)} Z_{*}(s)
\label{new:c2}
\end{align}

Putting value of $Z_{*}(s)$ from Eq. \eqref{eqn:125b} in  Eq. \eqref{new:c2} we get:

\begin{align}
& C_{2} (s) = \frac {(k_1 \alpha_1 P )\beta_1 Z_T} {[(s+\alpha_2+k_2)(s+\alpha_1P)-(\alpha_1\alpha_2P)](s + \beta_2 + k_1)} N_{L,R}(s) \nonumber \\  & C_{2} (s) =   \frac {(k_1 \alpha_1 P )\beta_1 Z_T} {[(s+\alpha_2+k_2)(s+\alpha_1P)-(\alpha_1\alpha_2P)](s + \beta_2 + k_1)} {\mathds 1}_R^T (sI - H)^{-1} {\mathds 1}_T U(s)
\label{eqn:c2}
\end{align}

%


%

The final result for ERC-OM link with RC receiver circuit is obtained by putting the values of $C_1(s)$ and $C_2(s)$ Eq. \eqref{eqn:3b5a} as follows:

\begin{align}
& N_{X} (s)   = r[\frac {(k_1 C_1(s)+ \alpha_2 C_2(s)/(1+r)} {s + \alpha_1 P_T/(1+r)}]  \\
& N_{X} (s)   =  r \frac{k_1 \beta_1 Z_T}{(s + \beta_2 + k_1)} [\frac {( 1 + \alpha_1P_T / [(s+\alpha_2+k_2)(s+\alpha_1P_T)-(\alpha_1\alpha_2P_T)]) / (1+r)}{s + \alpha_1 P_T/(1+r)}]  {\mathds 1}_R^T (sI - H)^{-1}{\mathds 1}_T U(s)
\label{eqn:3b6a}
\end{align}
Taking
\begin{align}
& Q(s) = {r [\frac {( 1 + \alpha_1P_T / [(s+\alpha_2+k_2)(s+\alpha_1P_T)-(\alpha_1\alpha_2P_T)]) / (1+r)}{s + \alpha_1 P_T/(1+r)}]} {\mathds 1}_R^T (sI - H)^{-1}{\mathds 1}_T  
\label{eqn:3b7a}
\end{align}

We obtain final result as:

\begin{align}
& N_{X}(s)   =  \underbrace {Q(s)\frac{k_1 \beta_1 Z_T}{(s + \beta_2 + k_1)}} _{\tilde{\Psi}(s)} U(s)
\end{align}

Which is the same result as Equation \eqref{eqn:3b6}.

Note that we can obtain the final result for the ERC-OM link with CATREG circuit in the output module in similar way, however we are not presenting the complete derivation for this case. An important similarity is that the expression for $C_{1} (s)$ is same for both RC and CATREG circuits in the output module. However for CATREG circuit in the output module the expression for $C_{2} (s)$ is different which is obtained as:  
\begin{align}
 & C_{2} (s) =   \frac {(k_1 \alpha_1 P_T  )\beta_1 Z_T} {[(s+\alpha_2+k_2)(s+\alpha_1P_T + k_+ - rk_0)-(\alpha_1\alpha_2P_T )](s + \beta_2 + k_1)} N_{L,R} (s)
\label{eqn:3bdss}
\end{align}

}

\bibliographystyle{IEEEtran}
\bibliography{nano2017,book,hamdan_ref}
\ifCLASSOPTIONcaptionsoff
  \newpage
\fi

\end{document}